\documentclass[aps,prl,twocolumn,superscriptaddress,showpacs,showkeys,amsmath,amssymb]{revtex4}

\usepackage[T1]{fontenc}
\usepackage[utf8]{inputenc}
\usepackage{textcomp}

\usepackage{float}
\usepackage{color}
\usepackage{bm}
\usepackage{amsmath}
\usepackage{amssymb}
\usepackage{graphicx}

\makeatletter

\usepackage{latexsym}
\usepackage[ngerman,english]{babel}

\begin{document}

\title{Thermodynamics of the pyrochlore Heisenberg ferromagnet with arbitrary spin $S$}

\author{Patrick M\"{u}ller}
\affiliation{Institut f\"{u}r theoretische Physik,
          Otto-von-Guericke-Universit\"{a}t Magdeburg,
          P.O. Box 4120, 39016 Magdeburg, Germany}

\author{Andre Lohmann}
\affiliation{Institut f\"{u}r theoretische Physik,
          Otto-von-Guericke-Universit\"{a}t Magdeburg,
          P.O. Box 4120, 39016 Magdeburg, Germany}

\author{Johannes Richter}
\affiliation{Institut f\"{u}r theoretische Physik,
          Otto-von-Guericke-Universit\"{a}t Magdeburg,
          P.O. Box 4120, 39016 Magdeburg, Germany}

\author{Oleg Menchyshyn}
\affiliation{Institute for Condensed Matter Physics,
          National Academy of Sciences of Ukraine,
          Svientsitskii Street 1, 79011 L'viv, Ukraine}

\author{Oleg Derzhko}
\affiliation{Institute for Condensed Matter Physics,
          National Academy of Sciences of Ukraine,
          Svientsitskii Street 1, 79011 L'viv, Ukraine}
\affiliation{Institut f\"{u}r theoretische Physik,
          Otto-von-Guericke-Universit\"{a}t Magdeburg,
          P.O. Box 4120, 39016 Magdeburg, Germany}
\affiliation{Department for Theoretical Physics,
          Ivan Franko National University of L'viv,
          Drahomanov Street 12, 79005 L'viv, Ukraine}
\affiliation{Abdus Salam International Centre for Theoretical Physics,
          Strada Costiera 11, 34151 Trieste, Italy}

\date{\today}

\begin{abstract}
We use the rotation-invariant Green's function method (RGM) and the high-temperature expansion (HTE) 
to study the thermodynamic properties of the spin-$S$ Heisenberg ferromagnet on the pyrochlore lattice. 
We examine the excitation spectra as well as various thermodynamic quantities, 
such as the order parameter (magnetization), the uniform static susceptibility, the correlation length, the spin-spin correlations, and the specific heat,
as well as the static and dynamic structure factors. 
We discuss the influence of the spin quantum number $S$ on the temperature dependence of these quantities.
We compare our results for the pyrochlore ferromagnet with the corresponding ones for the simple-cubic lattice 
both having the same coordination number $z=6$. 
We find a significant suppression of magnetic ordering for the pyrochlore lattice 
due to its geometry with corner-sharing tetrahedra.
\end{abstract}

\pacs{
75.10.-b, 
75.10.Jm  
}

\keywords{spin-$S$ Heisenberg ferromagnet, 
pyrochlore lattice, 
rotation-invariant Green's function method, 
high-temperature expansion}

\maketitle

\section{Introduction}
\label{sec1}

There has been much interest in frustrated spin systems during the last decades
\cite{LNP2004,Lacroix2011}.
Competing interactions due to lattice geometry 
together with quantum fluctuations due to small coordination numbers and/or low spin quantum numbers $S$ 
can prevent magnetic ordering even in the ground state 
and give rise to a rich diversity of quantum phases.
The most popular lattices used for the study of frustrations are the lattices of corner-sharing triangles or tetrahedra.
In particular, 
the network of corner-sharing tetrahedra known as the three-dimensional pyrochlore lattice 
was in the focus of many researchers during the past 25 years both from experimental and theoretical sides \cite{Gardner2010}.

Among the magnetic models on the pyrochlore lattice the quantum Heisenberg antiferromagnet is likely the most challenging one
\cite{Harris1991,Canals1998,Isoda1998,Koga2001,Tsunetsugu2001,Berg2003,Moessner2006,Henley2006,Normand2014}.
Thus, until now  neither the nature of the ground state is understood nor precise values for the ground state energy are available.
On the material side, 
there are numerous realizations of antiferromagnetically coupled Heisenberg spins on the pyrochlore lattice \cite{Gardner2010},
however, side effects, such as magnetostatic dipole-dipole interactions or coupling to lattice degrees of freedom, 
may influence the magnetic properties of pyrochlore compounds.

Much less attention has been payed to the quantum pyrochlore {\em ferromagnet}.
Clearly, the ground state of the ferromagnet and its energy are not affected by geometrical frustration. 
It is also clear, 
that the set of the eigenstates of the Heisenberg Hamiltonian does not depend on the sign of the exchange interaction,
but the arrangement of eigenstates according to their energy is opposite for antiferromagnetic and ferromagnetic interactions, 
i.e., the low-energy states of the antiferromagnet correspond to the high-energy states of the ferromagnet.
Therefore, 
for the ferromagnet the frustrated geometry of the corner-sharing tetrahedra
leads to a shift of the upper bound of the spectrum 
(given by the absolute value of the antiferromagnetic ground-state energy) 
towards the (unshifted) ferromagnetic ground-state energy.
Thus, due to frustration the energy spectrum becomes ``compressed'' and, as a result, 
the excited states for the ferromagnet on a frustrated lattice become  easier accessible as the temperature increases.
This {\em finite-temperature frustration effect} in ferromagnets manifests itself in a decrease of the Curie temperature $T_c$ \cite{Schmalfus2005,Lohmann2014,Mueller2015}.  
With respect to the pyrochlore ferromagnet 
it is reasonable to compare it with the corresponding ferromagnet on the bipartite simple-cubic lattice,
where  no frustration effects are present. 
Since the coordination number for both lattices is the same, $z=6$, 
the thermodynamics on the mean-field level of both models is identical.
However, using more accurate approaches 
the influence of the lattice geometry should be visible in the temperature profile of thermodynamic quantities.

There are only a few universal approaches 
to calculate thermodynamic quantities of Heisenberg quantum spin systems of arbitrary lattice geometry,
such as  
the Green-function technique \cite{Tyablikov1967,Gasser2001,Froebrich2006} 
and 
the high-temperature expansion \cite{elstner1993,2DJ1J2,singh2012,Kapellasite,Oitmaa2006,Bernu2001,Lohmann2011,Lohmann2014,Richter2015,Schmidt2017};
both are used in the present study 
to derive various finite-temperature characteristics of the pyrochlore Heisenberg ferromagnet with spin quantum number $S\ge 1/2$.

It is in order to mention here a solid-state realization  of the $S=1/2$ Heisenberg ferromagnet on the pyrochlore lattice \cite{Zhou2008,Onose2010,Mena2014}.
Lu$_2$V$_2$O$_7$ is a ferromagnetic, small-gap Mott insulator, that crystallizes in the pyrochlore structure,
i.e., the V$^{4+}$ ions carrying $S=1/2$ occupy the sites of the pyrochlore lattice.
However, 
the low symmetry of the pyrochlore lattice allows for a Dzyaloshinskii-Moriya interaction.
From the bulk measurements for Lu$_2$V$_2$O$_7$ it is known that the Curie temperature is $T_c=70$~K,
and neutron inelastic scattering data are in an excellent agreement with a minimal model 
that includes a nearest-neighbor Heisenberg exchange $\vert J\vert=8.22(2)$~meV 
and (possibly) a Dzyaloshinskii-Moriya interaction $D=1.5(1)$~meV, 
i.e., $T_c \approx 0.73\vert J\vert$. 
We will discuss the relation to our work in the summary section.

What follows  is organized as follows.
First we introduce the model (Sec.~\ref{sec2}) and explain the methods to be used (Sec.~\ref{sec3}).
Then we discuss the obtained results comparing the outcomes of two different methods, 
the rotation-invariant Green's function method and the high-temperature expansion, 
and the results for the pyrochlore and simple-cubic lattices (Sec.~\ref{sec4}).
We end up with conclusions emphasizing the peculiarities of the studied thermodynamics due to lattice geometry (Sec.~\ref{sec5}).

\section{Model}
\label{sec2}

\begin{figure}
\begin{center}
\includegraphics[clip=on,width=80mm,angle=0]{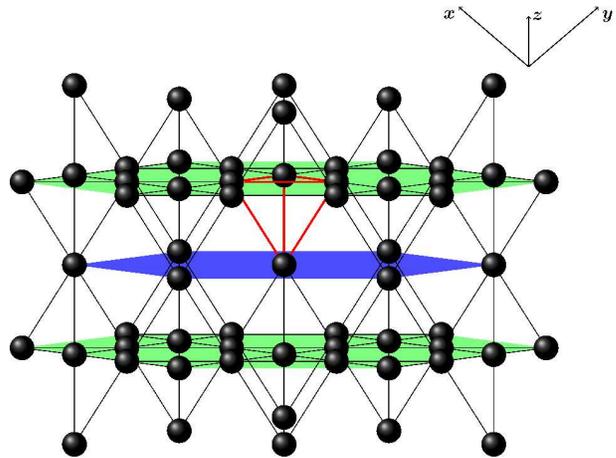}
\caption
{The pyrochlore lattice can be visualized as a structure which consists of alternating kagome and triangular planar layers. 
The kagome (triangular) planes are colored in green (blue).
The four-site unit cell is marked with the red bonds.}
\label{fig01} 
\end{center}
\end{figure}

We consider the Heisenberg model
\begin{eqnarray}
\label{01}
\hat{H}  = J\sum_{\langle m\alpha,n\beta\rangle} \hat{{\bm{S}}}_{m\alpha}\cdot\hat{{\bm{S}}}_{n\beta}
\end{eqnarray}
on the pyrochlore lattice, see Fig.~\ref{fig01}. 
The ferromagnetic nearest-neighbor coupling is set to $J=-1$ 
and arbitrary spin quantum number $S$ is considered, $\hat{{\bm{S}}}_{m\alpha}^2 = S(S+1)$. 
In the sum over all nearest-neighbor bonds in Eq.~(\ref{01}), 
the Latin indices denote the corresponding unit cell, whereas the Greek indices mark the corresponding spin within a unit cell,
see below.

For the presentation of the methods used in the present paper as well as for the discussion of the results 
it is useful to provide a short description of the pyrochlore lattice. 
The lattice can be visualized in different ways.
It can be described as four interpenetrating face-centered-cubic sublattices.
The edge length of the cubic cell of each face-centered-cubic sublattice is set to unity.
The origins of the four face-centered-cubic sublattices are taken to be
${\bf{r}}_1=(0,0,0)$,
${\bf{r}}_2=(0,1/4,1/4)$,
${\bf{r}}_3=(1/4,0,1/4)$,
and
${\bf{r}}_4=(1/4,1/4,0)$.
The sites of the face-centered-cubic lattice are determined by
${\bf{R}}_m=m_1{\bf{e}}_1 +m_2{\bf{e}}_2+m_3{\bf{e}}_3$,
where $m_1$, $m_2$, $m_3$ are integers 
and 
${\bf{e}}_1=(0,1/2,1/2)$,
${\bf{e}}_2=(1/2,0,1/2)$,
${\bf{e}}_3=(1/2,1/2,0)$.
Then for the sites of the pyrochlore lattice $m\alpha$, $m=1,\ldots,{\cal{N}}$, ${\cal{N}}=N/4$ 
we have ${\bf{R}}_{m\alpha}={\bf{R}}_m+{\bf{r}}_\alpha$,  
where $\alpha=1,2,3,4$ labels the sites in a unit cell.  
Geometrically this unit cell is a tetrahedron, where the corners are connected by $J$-bonds,
see the tetrahedron with red edges in Fig.~\ref{fig01}.
The distance between the nearest-neighbor sites is $1/\sqrt{8}$,
the distance between the next-nearest-neighbor sites is $\sqrt{3/8}$ etc.
The pyrochlore lattice can be also viewed as alternating planes of triangular and kagome lattices, 
see Fig.~\ref{fig01}.
Each spin on the pyrochlore lattice has $z=6$ nearest neighbors.
Thus the comparison with the simple-cubic lattice with the same coordination number $z=6$ is natural.

\section{Methods}
\label{sec3}

\subsection{Rotation-invariant Green's function method (RGM)}
\label{sec3A}

Double-time temperature-dependent Green's functions are widely used in quantum many-body physics \cite{Tyablikov1967,Gasser2001,Froebrich2006}.
An important contribution to the development of this technique was made by Kondo and Yamaji \cite{Kondo1972}.
They considered the hierarchy of the equations of motion of the Green's functions 
for the one-dimensional  $S = 1/2$ Heisenberg model.
In order to describe short-range order at $T>0$  
they decoupled the hierarchy at one-step further than Tyablikov's decoupling 
(also called random-phase approximation (RPA)) \cite{Tyablikov1967,Gasser2001,Mi2016}
and established  rotational invariance by setting $\langle \hat{S}^z_{i}\rangle=0$ in the equations of motions. 
In addition, the approximate decoupling of higher-order correlation functions is partly ``repaired'' 
by introducing so-called vertex parameters. 
Within this rotation-invariant Green's function method (RGM)
magnetic long-range order is then described by the long-range term in the spin-spin correlation function.
Over time the RGM was further developed and brought into shape
to include arbitrary quantum spin numbers $S$ in higher-dimensional lattices with non-primitive unit cells
\cite{Rhodes1973,Shimahara1991,Suzuki1994,barabanov94,Ihle1997,Ihle1999,Yu2000,Ihle2001,Bernhard2002,Schmalfus2004,Junger2005,Haertel2010,Antsygina2008,Mikheyenkov2016,Vladimirov2017}.
Nowadays the RGM is a well established method 
and has been the tool of choice in numerous recent publications 
on the theory of spin systems including geometrically frustrated ones
\cite{Yu2000,Ihle2001,Bernhard2002,Schmalfus2004,Schmalfus2005,Haertel2010,Mueller2015,Junger2005,Antsygina2008,Mikheyenkov2016,Vladimirov2017}.

The key point of the double-time temperature-dependent Green's functions approach 
is the determination of a set of Green's functions
$\langle\langle\hat{S}_{\mathbf{q}\alpha}^{\mu};\hat{S}_{\mathbf{\mathbf{q}\beta}}^{\nu}\rangle\rangle_{\omega}$
which are related to the dynamic susceptibilities of the spin system by 
$\langle\langle\hat{S}_{\mathbf{q}\alpha}^{\mu};\hat{S}_{\mathbf{\mathbf{q}\beta}}^{\nu}\rangle\rangle_{\omega}
=
-\chi^{\mu\nu}_{\mathbf{q}\alpha\beta}(\omega)$ \cite{Tyablikov1967,Gasser2001}.
Here typically $\mu\nu$ is $+-$ or $zz$
and
$\hat{S}_{\mathbf{q}\alpha}^{+}=(1/\sqrt{{\cal{N}}})\sum_{m}\exp(-{\rm{i}}{\bf{q}}\cdot{\bf{R}}_m) \hat{S}_{m\alpha}^{+}$
etc.,
where the sum runs over all unit cells,
$m=1,\ldots,{\cal{N}}$, ${\cal{N}}=N/4$.
Furthermore,
$\langle\langle \hat{X};\hat{Y}\rangle\rangle=-{\rm{i}}\Theta(t-t^\prime)\langle[\hat{X}(t),\hat{Y}(t^\prime)]_-\rangle$
and the subscript $\omega$ means the Fourier-transform with respect to the time $t-t^\prime$.
The Green's functions obey a set of equations of motion, 
which involves Green's functions of higher order than the initial ones.
The RGM considers the equation of motion up to the second order,
i.e.,
\begin{eqnarray}
\label{02}
\omega^{2}\langle\langle\hat{S}_{\mathbf{q}\alpha}^{z};\hat{S}_{\mathbf{\mathbf{q}\beta}}^{z}\rangle\rangle_{\omega}
=
\langle[\textrm{i}\dot{\hat{S}}_{\mathbf{q}\alpha}^{z},\hat{S}_{\mathbf{\mathbf{q}\beta}}^{z}]_{-}\rangle
-\langle\langle\ddot{\hat{S}}_{\mathbf{q}\alpha}^{z};\hat{S}_{\mathbf{\mathbf{q}\beta}}^{z}\rangle\rangle_{\omega}.
\end{eqnarray}
The operator 
$-\ddot{\hat{S}}_{\mathbf{q}\alpha}^{z}=[[{\hat{S}}_{\mathbf{q}\alpha}^z,\hat{H}]_-,\hat{H}]_-$ 
consists of several combinations of three-spin operators 
made of $\hat{S}_{\mathbf{q}\alpha}^{\mu}$ with $\mu=+,-,z$,
which can be obtained explicitly using the commutation relations
$[\hat{S}^x,\hat{S}^y]_-={\rm{i}}\hat{S}^z$ etc.
These products of three-spin operators have to be simplified by a decoupling scheme.
The spirit of the decoupling within $-\ddot{\hat{S}}_{\mathbf{q}\alpha}^{z}$ 
is exemplarily sketched as follows:
\begin{eqnarray}
\label{03}
\hat{S}_{A}^{+}\hat{S}_{B}^{-}\hat{S}_{C}^{z}&\rightarrow&\alpha^{}_{AB}c^{+-}_{AB}\hat{S}_{C}^{z},
\nonumber\\
\hat{S}_{A}^{+}\hat{S}_{A}^{-}\hat{S}_{B}^{z}&\rightarrow& \frac{2}{3}S(S+1)\hat{S}_{B}^{z},
\nonumber\\
\hat{S}_{A}^{+}\hat{S}_{B}^{-}\hat{S}_{B}^{z}&\rightarrow&\lambda^{}_{AB}c^{+-}_{AB}\hat{S}_{B}^{z}.
\end{eqnarray}
Here 
$A$, $B$, and $C$ represent different sites of the pyrochlore lattice, 
$c^{+-}_{AB}=\langle\hat{S}_{A}^{+}\hat{S}_{B}^{-}\rangle$,
and the conservation of total $S^z$ is implied, 
i.e., $c^{+z}_{AB}=c^{-z}_{AB}=0$.
In Eq.~(\ref{03}) two kinds of so-called vertex parameters $\alpha^{}_{AB}$ and $\lambda^{}_{AB}$ have been introduced 
to improve the approximation made by the  decoupling.
The vertex parameter $\alpha^{}_{AB}$ appears in the decoupling scheme if all sites are pairwise different,
see the first line in Eq.~(\ref{03}). 
In the second line in Eq.~(\ref{03})
the correlation $\langle \hat{S}_{A}^{+}\hat{S}_{A}^{-}\rangle$ is determined 
by the operator identity $\hat{\mathbf{S}}^2 = \hat{S}^+\hat{S}^- - \hat{S}^z + (\hat{S}^z)^2$.
The vertex parameter $\lambda^{}_{AB}$ introduced in the third line of Eq.~(\ref{03}) 
appears only for $S>1/2$ if two site indices coincide 
and the remaining correlation function cannot be determined by an operator identity.

After implementation of this approximation,
the set of equations in Eq.~(\ref{02}) can be compactly written in the matrix form as follows: 
\begin{eqnarray}
\label{04}
(\omega^2 I - F_{\mathbf{q}})\chi^{{+-}}_{\mathbf{q}}(\omega) = -M_{\mathbf{q}}.
\end{eqnarray}
Here $I$ denotes the  $4\times 4$  unit matrix
and
we have introduced the Hermitian $4\times 4$ matrices
$F_\mathbf{q}$ (the frequency matrix), 
$\chi^{{+-}}_\mathbf{q}(\omega)$ (the susceptibility matrix),
and
$M_\mathbf{q}$ (the momentum matrix).
Clearly, the $4\times 4$ matrices appear here because the unit cell contains four sites.
For the matrix elements of the momentum matrix and the frequency matrix for the model at hand explicit expressions can be found:
\begin{widetext}
\begin{eqnarray}
\label{05}
\frac{M_{\mathbf{q}11}}{J}  = \frac{M_{\mathbf{q}22}}{J} = \frac{M_{\mathbf{q}33}}{J} = \frac{M_{\mathbf{q}44}}{J} = -12c_{100}, 
\nonumber\\
\frac{M_{\mathbf{q}12}}{J}  = 4c_{100}\textrm{cos}\frac{q_{x}+q_{y}}{4},
\;\;\;
\frac{M_{\mathbf{q}13}}{J}  =  4c_{100}\textrm{cos}\frac{q_{x}+q_{z}}{4},
\;\;\;
\frac{M_{\mathbf{q}14}}{J}  =  4c_{100}\textrm{cos}\frac{q_{y}+q_{z}}{4},
\nonumber\\
\frac{M_{\mathbf{q}23}}{J}  =  4c_{100}\textrm{cos}\frac{q_{y}-q_{z}}{4},
\;\;\;
\frac{M_{\mathbf{q}24}}{J}  =  4c_{100}\textrm{cos}\frac{q_{x}-q_{z}}{4},
\;\;\;
\frac{M_{\mathbf{q}34}}{J}  =  4c_{100}\textrm{cos}\frac{q_{x}-q_{y}}{4} 
\end{eqnarray}
and
\begin{eqnarray}
\frac{F_{\mathbf{q}11}}{J^{2}}  =  2\left(f_{1}+\tilde{\alpha}_{100}
\left(\cos\frac{q_{x}+q_{y}}{2}+\cos\frac{q_{x}+q_{z}}{2}+\cos\frac{q_{y}+q_{z}}{2}\right)\right),
\nonumber\\
\frac{F_{\mathbf{q}22}}{J^{2}}  =  2\left(f_{1} +\tilde{\alpha}_{100}
\left(\cos\frac{q_{x}+q_{y}}{2}+\cos\frac{q_{x}-q_{z}}{2}+\cos\frac{q_{y}-q_{z}}{2}\right)\right),
\nonumber\\
\frac{F_{\mathbf{q}33}}{J^{2}}  =  2\left(f_{1}+\tilde{\alpha}_{100}
\left(\cos\frac{q_{x}-q_{y}}{2}+\cos\frac{q_{x}+q_{z}}{2}+\cos\frac{q_{y}-q_{z}}{2}\right)\right),
\nonumber\\
\frac{F_{\mathbf{q}44}}{J^{2}}  =  2\left(f_{1}+\tilde{\alpha}_{100}
\left(\cos\frac{q_{x}-q_{y}}{2}+\cos\frac{q_{x}-q_{z}}{2}+\cos\frac{q_{y}+q_{z}}{2}\right)\right),
\nonumber
\end{eqnarray}
\begin{eqnarray}
\label{06}
\frac{F_{\mathbf{q}12}}{J^{2}}  =  \frac{2}{3}\left(6\tilde{\alpha}_{100}\cos\frac{q_{z}}{2}\cos\frac{q_{x}-q_{y}}{4}-f_{2}\cos\frac{q_{x}+q_{y}}{4}\right),
\nonumber\\
\frac{F_{\mathbf{q}13}}{J^{2}}  =  \frac{2}{3}\left(6\tilde{\alpha}_{100}\cos\frac{q_{y}}{2}\cos\frac{q_{x}-q_{z}}{4}-f_{2}\cos\frac{q_{x}+q_{z}}{4}\right),
\nonumber\\
\frac{F_{\mathbf{q}14}}{J^{2}}  =  \frac{2}{3}\left(6\tilde{\alpha}_{100}\cos\frac{q_{x}}{2}\cos\frac{q_{y}-q_{z}}{4}-f_{2}\cos\frac{q_{y}+q_{z}}{4}\right),
\nonumber\\
\frac{F_{\mathbf{q}23}}{J^{2}}  =  \frac{2}{3}\left(6\tilde{\alpha}_{100}\cos\frac{q_{x}}{2}\cos\frac{q_{y}+q_{z}}{4}-f_{2}\cos\frac{q_{y}-q_{z}}{4}\right),
\nonumber\\
\frac{F_{\mathbf{q}24}}{J^{2}}  =  \frac{2}{3}\left(6\tilde{\alpha}_{100}\cos\frac{q_{y}}{2}\cos\frac{q_{x}+q_{z}}{4}-f_{2}\cos\frac{q_{x}-q_{z}}{4}\right),
\nonumber\\
\frac{F_{\mathbf{q}34}}{J^{2}}  =  \frac{2}{3}\left(6\tilde{\alpha}_{100}\cos\frac{q_{z}}{2}\cos\frac{q_{x}+q_{y}}{4}-f_{2}\cos\frac{q_{x}-q_{y}}{4}\right)
\end{eqnarray}
\end{widetext}
with
$f_{1}= 2S(S+1) + 3(\tilde{\lambda}_{100} + 2(\tilde{\alpha}_{100}+\tilde{\alpha}_{110})+\tilde{\alpha}_{200})$,
$f_{2}= 2S(S+1) + 3(\tilde{\lambda}_{100}+5\tilde{\alpha}_{100}+2\tilde{\alpha}_{110}+\tilde{\alpha}_{200})$,
$\tilde{\lambda}_{ijk}=\lambda_{ijk}c_{ijk}$,
and 
$\tilde{\alpha}_{ijk}=\alpha_{ijk}c_{ijk}$.
Here the indices $ijk$ correspond to the vector $\mathbf{R}=i\mathbf{r}_2+j\mathbf{r}_3+k\mathbf{r}_4$, 
i.e., 
$c_{ijk} \equiv \langle \hat{S}^+_{\mathbf{0}} \hat{S}^-_{\mathbf{R}}\rangle$. 
Note also that on grounds of the lattice symmetry 
the set of non-equivalent correlators has been reduced in Eqs.~(\ref{05}) and  (\ref{06}).

The derivation of Eqs.~(\ref{04}), (\ref{05}), and (\ref{06}) 
is the central task within the RGM approach.
To perform the underlying tedious calculations,
we use the symbolic computation software {\it Mathematica}.
We notice that Eqs.~(\ref{04}), (\ref{05}), and (\ref{06}) hold for antiferromagnetic coupling $J=1$, too.
For easy references, 
we provide in addition the corresponding equations for the $S=1/2$ simple-cubic Heisenberg model in Appendix~A 
(see also Refs.~\cite{Kawabe1973,preprint}).

Going back to Eq.~(\ref{04}),
it is important to note that the momentum matrix $M_\mathbf{q}$ and the frequency matrix $F_\mathbf{q}$ commute:
$[M_\mathbf{q},F_\mathbf{q}]_- =0$.
Let us denote as $\vert{\gamma\mathbf{q}}\rangle$, $\gamma=1,2,3,4$
the common eigenvectors of the matrices $M_\mathbf{q}$ and $F_\mathbf{q}$.
Moreover, let us introduce their eigenvalues,
i.e.,
$M_{\mathbf{q}}|{\gamma\mathbf{q}}\rangle=m_{\gamma\mathbf{q}}|{\gamma\mathbf{q}}\rangle$
and
$F_{\mathbf{q}}|{\gamma\mathbf{q}}\rangle=\omega^2_{\gamma\mathbf{q}}|{\gamma\mathbf{q}}\rangle$.
As usually,
the square root of the eigenvalues $\omega^2_{\gamma\mathbf{q}}$ yields the branches of the excitation spectrum $\omega_{\gamma\mathbf{q}}$, $\gamma=1,2,3,4$. 

Before finding $\chi^{{+-}}_{\mathbf{q}\alpha\beta}(\omega)$ from Eq.~(\ref{04}),
it is worth to discuss the eigenvalues of the matrices $M_\mathbf{q}$ and $F_\mathbf{q}$,
that is, $m_{\gamma\mathbf{q}}$ and $\omega^2_{\gamma\mathbf{q}}$, respectively.
We have found
\begin{eqnarray}
\label{07} 
\frac{m_{1\mathbf{q}}}{J} = \frac{m_{2\mathbf{q}}}{J} = \frac{m_{3\mathbf{q}}}{J}+\frac{m_{4\mathbf{q}}}{J} & = & -16 c_{100},
\nonumber\\
\frac{m_{3\mathbf{q}}}{J}-\frac{m_{4\mathbf{q}}}{J} & = & -8 c_{100}D_{\mathbf{q}}
\end{eqnarray}
with
\begin{eqnarray}
\label{08}
D_{\mathbf{q}}^2  =  1 +\cos\frac{q_x}{2}\cos\frac{q_y}{2} +\cos\frac{q_x}{2}\cos\frac{q_z}{2} +\cos\frac{q_y}{2}\cos\frac{q_z}{2}
\end{eqnarray}
and
\begin{eqnarray}
\label{09} 
\frac{\omega^2_{1\mathbf{q}}}{J^2} =\frac{\omega^2_{2\mathbf{q}}}{J^2}
&=& \frac{8}{3}(2S(S+1)+3\tilde{\lambda}_{100}
\nonumber\\
&+& 9\tilde{\alpha}_{100} + 6 \tilde{\alpha}_{110}+3 \tilde{\alpha}_{200}),
\nonumber\\
\frac{\omega^2_{3\mathbf{q}}}{J^2}+\frac{\omega^2_{4\mathbf{q}}}{J^2} 
&=& \frac{8}{3}(2S(S+1)+3\tilde{\lambda}_{100}
\nonumber\\
&+& 3(D_{\mathbf{q}}^2-1)\tilde{\alpha}_{100}+6\tilde{\alpha}_{110}+3\tilde{\alpha}_{200}),
\nonumber\\
\frac{\omega^2_{3\mathbf{q}}}{J^2}-\frac{\omega^2_{4\mathbf{q}}}{J^2} 
&=& \frac{8}{3}D_{\mathbf{q}}S(S+1)
\nonumber\\
&+& 4 D_{\mathbf{q}} (\tilde{\lambda}_{100}+3\tilde{\alpha}_{100}+2\tilde{\alpha}_{110}+\tilde{\alpha}_{200}).
\end{eqnarray}
As it immediately follows from Eq.~(\ref{09}),
there are two dispersionless (flat) branches of the spectrum, 
i.e., $\omega_{1\mathbf{q}}$ and $\omega_{2\mathbf{q}}$ do not depend on $\mathbf{q}$.
We may also consider the limit $|\mathbf{q}|\rightarrow 0^+$
when
$D_{\mathbf{q}}^2\rightarrow4$, 
$m_{3\mathbf{q}}\rightarrow m_{1\mathbf{q}}=m_{2\mathbf{q}}$, 
$m_{4\mathbf{q}}\rightarrow 0^+$, 
$\omega^2_{3\mathbf{q}}\rightarrow\omega^2_{1\mathbf{q}}=\omega^2_{2\mathbf{q}}$,
and 
$\omega^2_{4\mathbf{q}}\rightarrow 0^+$.
Evidently, 
$\omega_{4\mathbf{q}}$ is the acoustic branch of the spectrum. 
It is obvious, that the excitation energies calculated within the RGM, see Eq.~(\ref{09}), 
exhibit a temperature renormalization that is proportional to the correlation functions. 
Moreover, the renormalization is wave-length dependent for the dispersive branches $\omega_{3{\bf{q}}}$ and $\omega_{4{\bf{q}}}$.
That is different to the RPA, 
where the temperature renormalization of the excitations is independent of the wavelength and proportional to the magnetization, 
see, e.g., Refs.~\cite{Tyablikov1967,Gasser2001},
i.e.,
the RPA fails in describing magnetic excitations (and also magnetic short-range order) for $T>T_c$.

\begin{figure}
\centering 
\includegraphics[clip=on,width=80mm,angle=0]{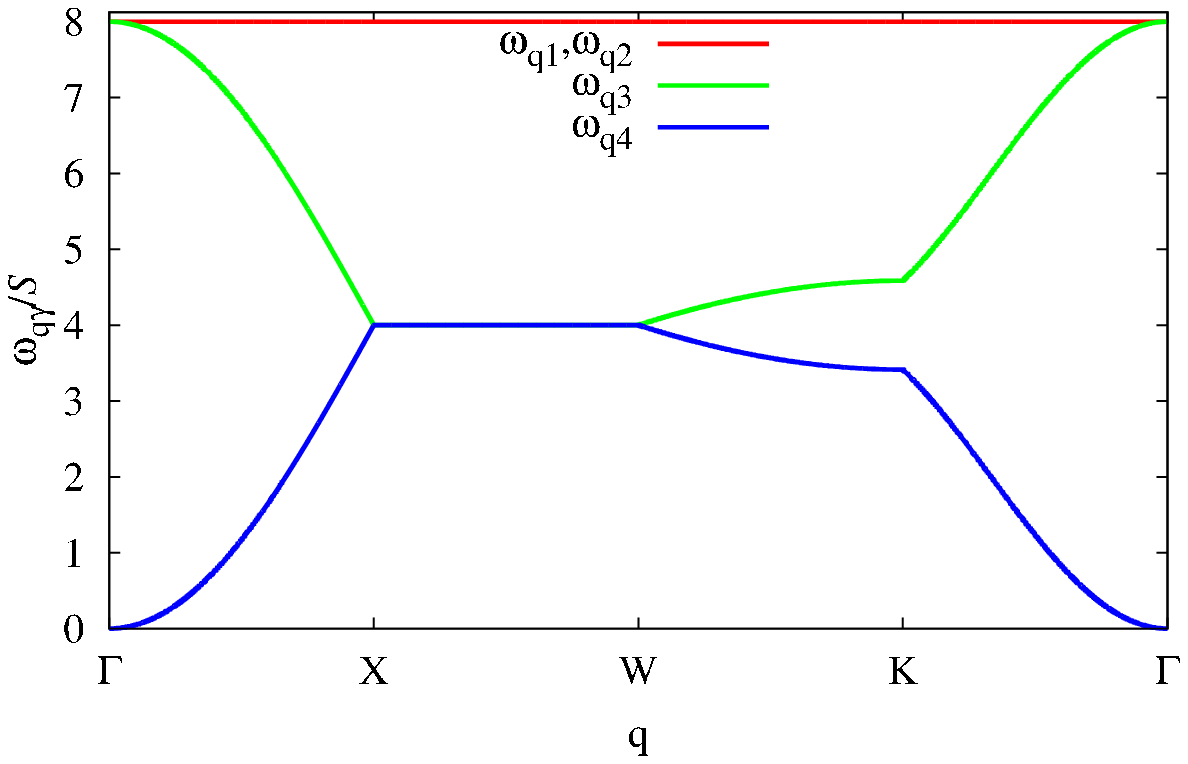}\\
\vspace{3mm}
\includegraphics[clip=on,width=80mm,angle=0]{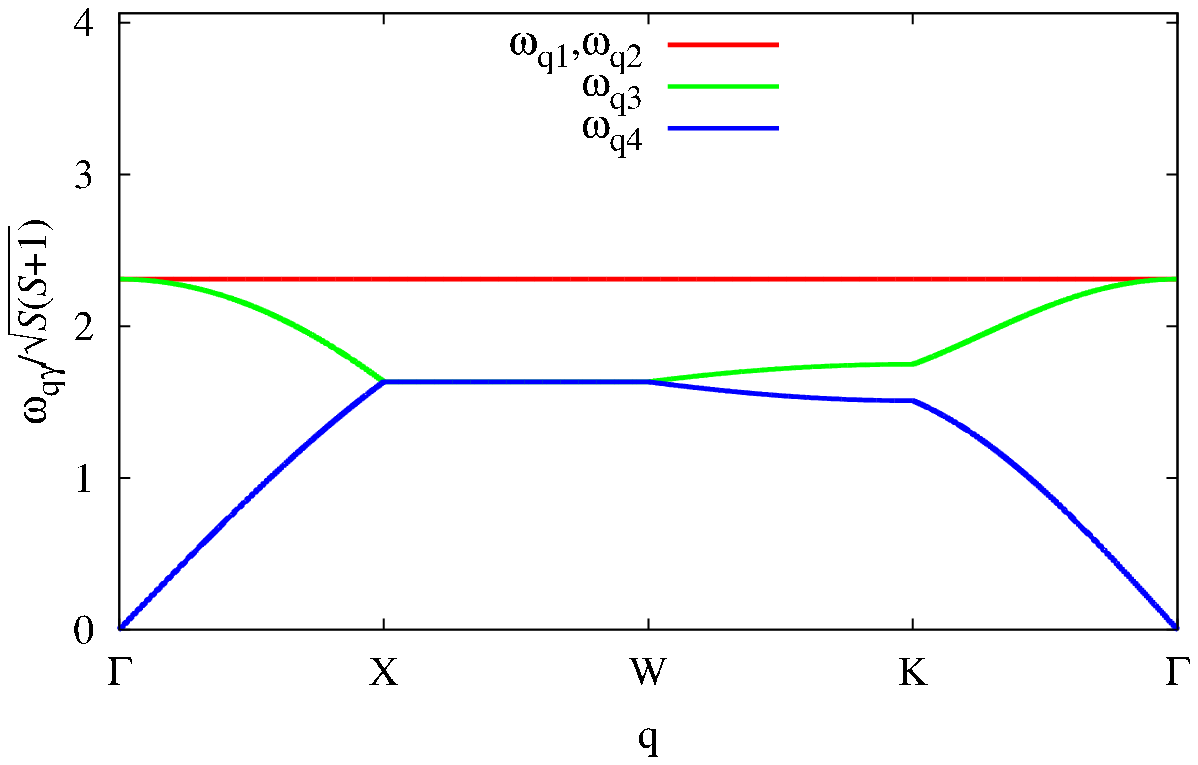}
\protect
\caption
{Dispersion of the excitation energies $\omega_{\mathbf{q}\gamma}$
(Eq.~(\ref{09}), $J=-1$)
at zero temperature $T=0$ (upper panel) and in the infinite-temperature limit $T\to\infty$ (lower panel). 
Note that $\omega_{\mathbf{q}\gamma}/S$ is independent of $S$ at $T=0$,
whereas $\omega_{\mathbf{q}\gamma}/\sqrt{S(S+1)}$ is independent of $S$ at $T\to\infty$.
The points $\Gamma$, X, W and K in the first Brillouin zone of a face-centered-cubic Bravais lattice 
are given by $\Gamma=(0,0,0)$, X$=(0,2\pi,0)$, W$=(\pi,2\pi,0)$, K$=(3\pi/2,3\pi/2,0)$, 
see, e.g., Ref.~\cite{symmetry_points}.}
\label{fig02} 
\end{figure}

At zero and infinite temperatures, 
we can get simplified expressions for the excitation energies given in Eq.~(\ref{09}). 
For $T=0$ we have $c_{ijk}=2S^2/3$, $\alpha_{ijk}=3/2$, and $\lambda_{ijk}=2-1/S$, see below.
As a result,
we get 
$\omega_{1{\bf{q}}}^2/J^2=\omega_{2{\bf{q}}}^2/J^2=64S^2$,
$\omega_{3{\bf{q}}}^2/J^2=4S^2(D_{\mathbf{q}}+ 2)^2$,
and
$\omega_{4{\bf{q}}}^2/J^2=4S^2(D_{\mathbf{q}}- 2)^2$.
As $T\to \infty$, we have $c_{ijk}=0$ resulting in 
$\omega_{1{\bf{q}}}^2/J^2=\omega_{2{\bf{q}}}^2/J^2=16S(S+1)/3$,
$\omega_{3{\bf{q}}}^2/J^2=4S(S+1)(2+D_{\mathbf{q}})/3$,
and
$\omega_{4{\bf{q}}}^2/J^2=4S(S+1)(2-D_{\mathbf{q}})/3$.
The branches of the spectrum (\ref{09}) in the ground state and in the infinite-temperature limit 
are shown in Fig.~\ref{fig02}.

Although the eigenvectors $\vert \gamma {\bf{q}}\rangle$ of the matrices $M_\mathbf{q}$ and $F_\mathbf{q}$ 
are also known explicitly,
they are too lengthy to be presented here (but they are given in Appendix~B).
However, 
at the $\Gamma$ point $\mathbf{q}=\mathbf{0}$ the eigenvectors $\vert \gamma {\bf{q}}\rangle$ have a very simple form:
\begin{eqnarray}
\label{10}
|{1\mathbf{0}}\rangle
=
\frac{1}{\sqrt{2}}
\left(
\begin{array}{c}
-1\\
0\\
0\\
1
\end{array}
\right),
\;\;\;
|{2\mathbf{0}}\rangle
=
\frac{1}{\sqrt{2}}
\left(
\begin{array}{c}
-1\\
0\\
1\\
0
\end{array}
\right),
\nonumber\\
|{3\mathbf{0}}\rangle
=
\frac{1}{\sqrt{2}}
\left(
\begin{array}{c}
-1\\
1\\
0\\
0
\end{array}
\right),
\;\;\;
|{4\mathbf{0}}\rangle
=
\frac{1}{2}
\left(
\begin{array}{c}
1\\
1\\
1\\
1
\end{array}
\right) .
\end{eqnarray}
Note that the eigenvectors $|{1\mathbf{0}}\rangle$, $|{2\mathbf{0}}\rangle$, and $|{3\mathbf{0}}\rangle$
correspond to the three-fold degenerate eigenvalue 
(either $m_{\gamma \mathbf{0}}$ or $\omega^2_{\gamma \mathbf{0}}$, $\gamma=1,2,3$)
and therefore any linear combination of $|{1\mathbf{0}}\rangle$, $|{2\mathbf{0}}\rangle$, and $|{3\mathbf{0}}\rangle$
given in Eq.~(\ref{10}) also belongs to a set of the eigenvectors at the $\Gamma$ point $\mathbf{q}=\mathbf{0}$.
Interestingly, the eigenvectors $\vert \gamma {\bf{q}}\rangle$ do not depend on the temperature, see Appendix~B.

Let us come back to Eq.~(\ref{04}).
The set of dynamic susceptibilities (and thus the set of Green's functions) is determined and given by
\begin{eqnarray}
\label{11}
\chi^{+-}_\mathbf{q\alpha\beta}(\omega) 
&=& -\sum_{\gamma} \frac{m_{\gamma\mathbf{q}}}{\omega^2-\omega^2_{\gamma\mathbf{q}}}\langle\alpha|{\gamma\mathbf{q}}\rangle\langle{\gamma\mathbf{q}}|\beta\rangle,
\end{eqnarray}
where $\langle\alpha|{\gamma\mathbf{q}}\rangle$ is the $\alpha$th component of the eigenvector $|{\gamma\mathbf{q}}\rangle$.
The correlation functions are obtained by applying the spectral theorem
\begin{eqnarray}
\label{12}
c_{m\alpha,n\beta} 
&=& \frac{1}{\mathcal{N}}\sum_{\mathbf{q}\ne\mathbf{Q}}c_{\mathbf{q}\alpha\beta}\cos(\mathbf{q}\cdot\mathbf{r}_{m\alpha,n\beta}) 
\nonumber \\
&+&\sum_{\mathbf{Q}}C_{\mathbf{Q}\alpha\beta}\cos(\mathbf{Q}\cdot\mathbf{r}_{m\alpha,n\beta})
\end{eqnarray}
with
\begin{eqnarray}
\label{13}
c_{\mathbf{q}\alpha\beta}
&=&\sum_{\gamma}\frac{m_{{\gamma}{\bf{q}}}}{2\omega_{\gamma\mathbf{q}}}(1+2n(\omega_{\gamma\mathbf{q}}))\langle\alpha|{\gamma\mathbf{q}}\rangle\langle{\gamma\mathbf{q}}|\beta\rangle,
\end{eqnarray}
where $\mathcal{N}=N/4$ is the number of unit cells, 
$n(\omega)=1/(\exp(\omega/T)-1)$ is the Bose-Einstein distribution function, 
and $C_{\mathbf{Q}\alpha\beta}$ is the so-called condensation term which is related to magnetic long-range order, 
see, e.g., Refs.~\onlinecite{Shimahara1991,Ihle1997,Ihle2001}. 
In our case (ferromagnet) only one condensation term at $\mathbf{Q}=\mathbf{0}$ is relevant, 
i.e., $C_{\mathbf{0}\alpha\beta}=C_{\mathbf{0}}$, 
and the total magnetization is given by the expression $M=\sqrt{3C_{\mathbf{0}}/2}$. 

We end up this subsection with some comments on the self-consistent solution 
of the equations for the correlation functions $c_{100}$, $c_{110}$, $c_{200}$, 
the condensation term $C_{\mathbf{0}}$,
and the vertex parameters.
We mention first that we adopt the so-called minimal version of RGM 
which is a well established approximation for ferromagnets, 
i.e., we use only one vertex parameter in each class $\alpha_{ijk}=\alpha$, $\lambda_{ijk}=\lambda$.
We begin with the high-temperature limit when $C_{\mathbf{0}}=0$ (paramagnetic phase).
We have three equations for $c_{100}$, $c_{110}$, $c_{200}$ which follow from Eq.~(\ref{12}),
as well as the equation
\begin{eqnarray}
\label{14}
\frac{2}{3}S(S+1)
=
\frac{1}{\mathcal{N}}\sum_{\mathbf{q}\ne\mathbf{0}}c_{\mathbf{q}\alpha\alpha}
\end{eqnarray}
(the sum rule $3c_{m\alpha,m\alpha}/2=S(S+1)$)
which also follows from Eq.~(\ref{12}).
Now only one missing equation, say to determine $\lambda$, is left.
A usual assumption is to treat the ratio $r(T)=(\lambda(T)-\lambda(\infty))/(\alpha(T)-\alpha(\infty))$ as temperature independent, 
see, e.g., Refs.~\cite{Haertel2010,Junger2005,Mueller2015}.
The values $\alpha(\infty)=1$ and $\lambda(\infty)=1-3/(4S(S+1))$ at $T\rightarrow\infty$ are known 
and can be verified by comparison with the high-temperature expansion, 
see, e.g., Ref.~\cite{Junger2005}.
The values $\alpha(0)=3/2$ and $\lambda(0)=2-1/S$ at $T=0$ are also exactly known, see below.
Now,
solving the system of equations numerically, 
we calculate the (static) uniform susceptibility 
$\chi_{\mathbf{0}}\equiv\chi^{zz}_{\mathbf{0}}=\chi^{+-}_{\mathbf{0}}/2$.
The uniform susceptibility $\chi_{\mathbf{0}}$ is given by the expression
\begin{eqnarray}
\label{15}
\chi_{\mathbf{0}}
=
\underset{(\mathbf{q},\omega)\rightarrow(\mathbf{0},0)}{\textrm{lim}} 
\frac{1}{4}\sum_{\alpha}
\sum_{\beta}\frac{\chi_{\mathbf{q}\alpha\beta}^{+-}(\omega)}{2}
\nonumber\\
=
\underset{(\mathbf{q},\omega)\rightarrow(\mathbf{0},0)}{\textrm{lim}} \frac{1}{8} \sum_{\alpha,\beta}\chi_{\mathbf{q}\alpha\beta}^{+-}(\omega)
=
\underset{\mathbf{q}\rightarrow\mathbf{0}}{\textrm{lim}}\left(\frac{m_{4\mathbf{q}}}{2\omega^2_{4\mathbf{q}}}+\ldots\right) 
\nonumber\\
=-\frac{3c_{100}}{\Delta},
\nonumber\\
\frac{\Delta}{J}
=
2S(S+1)+3\tilde{\lambda}_{100}
-15\tilde{\alpha}_{100}+6\tilde{\alpha}_{110}+3\tilde{\alpha}_{200}.
\end{eqnarray}
At the critical temperature $T_c$, when $\Delta=0$, 
the uniform susceptibility $\chi_{\mathbf{0}}$ diverges.
Moreover, $\Delta=0$ holds for all temperatures below $T_c$.
By using Eq.~(\ref{15}) this can be cast into
\begin{eqnarray}
\label{16}
2S(S+1)+3\tilde{\lambda}_{100} -15\tilde{\alpha}_{100}+6\tilde{\alpha}_{110}+3\tilde{\alpha}_{200}=0.
\end{eqnarray}
Therefore, for $0\le T<T_c$ (ferromagnetic phase) the formula (\ref{16}) provides one more equation,  
which is necessary to determine one more quantity, namely, the condensation term $C_{\mathbf{0}} \ne 0$.

In the fully polarized ferromagnetic ground state we have 
$\langle \hat{{\bf{S}}}_{\mathbf{0}} \cdot \hat{{\bf{S}}}_{\mathbf{R}}\rangle=S^2$, 
i.e., $c_{ijk}=2S^2/3$,
and as a result Eq.~(\ref{16}) becomes $2S(S+1)+2S^2\lambda -4S^2\alpha=0$.
Considering the sum rule,
i.e., $2S/3=(1/(4{\cal{N}}))\sum_{{\bf{q}}\ne 0}\sum_\gamma (m_{\gamma{\bf{q}}}/(2\omega_{\gamma{\bf{q}}}))$
including Eqs.~(\ref{07}) and (\ref{09}) at $T=0$
we get a second equation for $\alpha$ and $\lambda$. 
Combining both equations we derive  $\alpha(T=0)=3/2$ and $\lambda(T=0)=2-1/S$.

Knowing the dynamic susceptibilities or the Green's functions (\ref{11}) and the correlation functions (\ref{12}), (\ref{13}),
we can easily obtain the static uniform susceptibility $\chi_{\mathbf{0}}$, 
the critical (Curie) temperature $T_c$, 
the correlation length $\xi$, 
the magnetization $M$, 
and the specific heat $C_V$.
Furthermore, 
using Eq.~(\ref{13}) we can also obtain the static structure factor $S_{\mathbf{q}}=3S_{\mathbf{q}}^{+-}/2$,
$S_{\bf{q}}^{+-}=\sum_{\alpha,\beta}c_{\mathbf{q}\alpha\beta}/4$, 
cf. Eq.~(\ref{15}).
Bearing in mind a comparison of the RGM static structure factor with the results coming from high-temperature series, 
see Sec.~\ref{sec3B} and Eq.~(\ref{17}),
it is useful to note the following, see Ref.~\cite{Collins1970}.
First,
combining the Kramers-Kronig relation and the fluctuation-dissipation theorem 
we have
$\chi_{\bf{q}}^{+-}=(1/(2\pi))\int_{-\infty}^{\infty}{\rm{d}}\omega (1-e^{-\omega/T})S^{+-}_{\bf{q}}(\omega)/\omega$
with
$S^{+-}_{\bf{q}}(\omega)=\int_{-\infty}^{\infty}{\rm{d}}t e^{{\rm{i}}\omega t}S^{+-}_{\bf{q}}(t)$.
At high temperatures $T=1/\beta \to\infty$ this can be cast into
$\chi_{\bf{q}}^{+-}\approx\beta S^{+-}_{\bf{q}}(t=0)$,
i.e., $\chi_{\bf{q}}^{+-}\approx\beta S^{+-}_{\bf{q}}$.
Second, 
by comparison of Eq.~(\ref{13}) in the limit $(\mathbf{q},\omega)\rightarrow(\mathbf{0},0)$ and Eq.~(\ref{15})
one concludes that $\chi_{\mathbf{0}}^{+-}$ and $\beta S^{+-}_{\mathbf{0}}$ coincide in the whole paramagnetic region $T>T_c$.
Last but not least,
the dynamic structure factor $S^{zz}_{{\bf{q}}}(\omega)=S^{+-}_{{\bf{q}}}(\omega)/2$ 
follows from the fluctuation-dissipation theorem,
i.e.,
$S^{+-}_{{\bf{q}}}(\omega)=(2/(1-e^{-\omega/T}))\Im\chi^{+-}_{{\bf{q}}}(\omega)$,
$\chi^{+-}_{{\bf{q}}}(\omega)=\sum_{\alpha,\beta}\chi^{+-}_{{\bf{q}}\alpha\beta}(\omega)/4$.

\subsection{High-temperature expansion (HTE)}
\label{sec3B}

Another universal straightforward approach to calculate thermodynamic quantities of spin systems 
is the high-temperature expansion (HTE) \cite{Oitmaa2006}.
More specifically,
in this study we use the HTE program of Ref.~\cite{Lohmann2014} 
freely available at \verb"http://www.uni-magdeburg.de/jschulen/HTE/"
in an extended version up to eleventh order 
to compute the series of the susceptibility 
$\chi_{\mathbf{0}}=\sum_nc_n\beta^n$ and the specific heat $C_V=\sum_n d_n\beta^n$ 
with respect to the inverse temperature $\beta=1/T$. 
To extend the region of validity of the power series, 
Pad\'{e} approximants are a useful and well-established transformation.
These approximants are ratios of two polynomials of degree $m$ and $n$, 
$[m,n]=P_m(\beta)/Q_n(\beta)$,
constructed in such a way that they reproduce correctly $m+n$ terms in the power series.
Using the power series of the uniform susceptibility $\chi_{\mathbf{0}}$ 
the roots of the equation $Q_n(\beta)=0$ can provide an estimate of the critical temperature $T_c$.
Alternatively one can consider the ratio $q_n=c_n/c_{n-1}$. 
Assuming critical behavior, 
i.e., $\chi_{\mathbf{0}}\propto(T-T_c)^{-\gamma}$, where $\gamma$ is the critical exponent,
$T_c$ is given by a linear fit $\lim_{n\rightarrow\infty}q_n\propto T_c+(\gamma-1)T_c/n$, 
see, e.g., Refs.~\cite{yeomans,Lohmann2014}. 

Furthermore, 
the high-temperature series of $\langle \hat{{\bf{S}}}_i\cdot\hat{{\bf{S}}}_j\rangle$ are calculated 
up to ninth order of $\beta$ 
following the lines illustrated in Refs.~\cite{Lohmann2011,Lohmann2014}. 
Using the series of the correlation functions we determine the magnetic structure factor 
\begin{eqnarray}
\label{17}
S_{\mathbf{q}}
=
\frac{1}{N}\sum_{i,j}\langle \hat{{\bf{S}}}_i\cdot\hat{{\bf{S}}}_j\rangle\cos(\mathbf{q}\cdot(\mathbf{R}_{i}-\mathbf{R}_{j})),  
\end{eqnarray}
see, e.g., Ref.~\cite{Richter2015}.
Here $i$ and $j$ are the sites of the pyrochlore lattice labeled in Sec.~\ref{sec2} by  $m\alpha$.
Evidently, $S_{\mathbf{q}}=3S_{\mathbf{q}}^{+-}/2$.
Furthermore,
substituting ${\mathbf{q}}=0$ in Eq.~(\ref{17}) 
one gets $S_{\mathbf{0}}=3\langle \hat{S}^z \hat{S}^z\rangle/N$ with $\hat{S}^z=\sum_i\hat{S}^z_i$.
On the other hand, calculating the uniform susceptibility per site $\chi^{zz}$ from the partition function 
one arrives at $\chi^{zz}=\beta(\langle \hat{S}^z \hat{S}^z\rangle-\langle \hat{S}^z\rangle\langle \hat{S}^z\rangle)/N$.
As a result,
we have $3\chi^{zz}_{\mathbf{0}}=\beta S_{\mathbf{0}}$
in the paramagnetic region,
and this general relation holds also for the RGM, see the end of Sec.~\ref{sec3A}.

\section{Finite-temperature properties} 
\label{sec4}

\subsection{Excitation spectra, spin stiffness and excitation velocity}
\label{sec4A}

We begin with a discussion of the excitation-energy spectra for the spin-$S$ Heisenberg ferromagnet on the pyrochlore lattice.
The dispersion relations are given in Eq.~(\ref{09}).
For the zero-temperature case the excitation energies,
given by 
$\omega_{1{\bf{q}}}^2/J^2=\omega_{2{\bf{q}}}^2/J^2=64S^2$,
$\omega_{3{\bf{q}}}^2/J^2=4S^2(D_{\mathbf{q}}+ 2)^2$,
and
$\omega_{4{\bf{q}}}^2/J^2=4S^2(D_{\mathbf{q}}- 2)^2$,
are plotted in the upper panel of Fig.~\ref{fig02}.
In general,
the excitation spectra have not to coincide with the linear-spin-wave energies:
while the latter ones are temperature-independent harmonic oscillations around the classical ($S\to\infty$) ground state, 
the excitations calculated within the RGM approach depend on temperature-dependent correlation functions.
However, at zero temperature both approaches yield identical excitation energies 
(see Fig.~1(c) of Ref.~\cite{Mena2014} for the linear-spin-wave results), 
since the excitations are above the exact ferromagnetic ground state that does not exhibit quantum fluctuations, 
and, therefore, the low-temperature excitations (\ref{09}) are the linear spin waves.
Note further that the excitation energies given above 
coincide with the one-magnon excitation branches reported in Ref.~\cite{Zhitomirsky2007},
see Eq.~(11) of that paper.
For finite temperatures 
the excitation energies (\ref{09}) are renormalized 
due the temperature dependence of the correlation functions entering Eq.~(\ref{09}). 
In the infinite-temperature limit again we find simple expressions:
$\omega_{1\mathbf{q}}^{2}/J^2  = \omega_{2\mathbf{q}}^{2}/J^2 = 16S(S+1)/3$,
$\omega_{3\mathbf{q}}^{2}/J^2  =  4S(S+1)(2+D_{\mathbf{q}})/3$,
and
$\omega_{4\mathbf{q}}^{2}/J^2  =  4S(S+1)(2-D_{\mathbf{q}})/3$.
The graphical presentation of these expressions is given in the lower panel of Fig.~\ref{fig02}.
Except the temperature renormalization of the absolute values of the energies, 
the most relevant change is found in the long-wavelength behavior 
(i.e., around the $\Gamma$ point), 
where we have a linear dependence on $\vert{\bf{q}}\vert$ at $T\to \infty$ instead of the quadratic dispersion at $T=0$, 
cf. the lower and upper panels of Fig.~\ref{fig02}.

\begin{figure}
\centering 
\includegraphics[clip=on,width=80mm,angle=0]{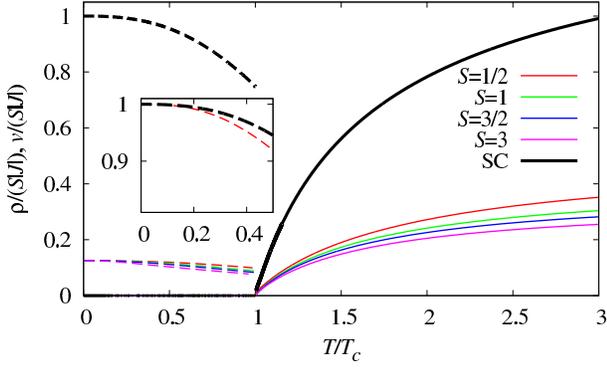}
\protect
\caption
{Main panel:
normalized spin stiffness $\rho/(S\vert J\vert)$ (dashed)
and 
normalized excitation velocity $v/(S\vert J\vert)$ (solid)  
as a function of the normalized temperature $T/T_c$.
We report results for different spin values $S=1/2,\,1,\,3/2,\,3$ 
for the pyrochlore-lattice case (thin lines) and for $S=1/2$ for the simple-cubic case (thick lines).
Inset:
spin stiffness $\rho/\rho(0)$ versus $T/\vert J\vert$
for the $S=1/2$ pyrochlore (dashed thin red line) and simple-cubic (dashed thick black line) lattices.}
\label{fig03} 
\end{figure}

Let us discuss the small-wavevector excitations in some detail.
By expansion around the $\Gamma=(0,0,0)$ point in the ${\bf{q}}$-space we get
\begin{eqnarray}
\label{18}
\omega^2_{4{\bf{q}}}
&\approx& 
v^2\vert{\bf{q}}\vert^2+\varrho^2 \vert{\bf{q}}\vert^4 
- \frac{J\Delta}{1152}\left(q_x^2q_y^2+q_x^2q_z^2+q_y^2q_z^2\right),
\nonumber\\
\frac{v^2}{J^2}
&=&
\frac{1}{24}(2S(S+1)+3\tilde{\lambda}_{100}
-15\tilde{\alpha}_{100}+6\tilde{\alpha}_{110}+3\tilde{\alpha}_{200})
\nonumber\\
&=&\frac{\Delta}{24 J},
\nonumber\\
\frac{\varrho^2}{J^2}
&=&
\frac{1}{4608}(-2S(S+1)-3\tilde{\lambda}_{100}
+87 \tilde{\alpha}_{100}-6 \tilde{\alpha}_{110}
\nonumber\\
&& -3 \tilde{\alpha}_{200})
=\frac{-\Delta + 72 \tilde{\alpha}_{100} J}{4608 J},
\end{eqnarray}
where $\Delta$ is defined in Eq.~(\ref{15}).
Clearly,
the excitation velocity vanishes below $T_c$ (where we have $\Delta=0$)
and the small-wavevector excitation energies depend quadratically on the wavevector 
with the spin stiffness $\rho=\varrho\vert_{\Delta=0}=\vert J\vert\sqrt{\tilde{\alpha}_{100}}/8$.
The stiffness is related to the stability of the ferromagnetic regime 
and can be an indicator of unusual effects like order-from-disorder effects or the rise of another magnetically ordered phase, 
see, e.g., Refs.~\cite{Mueller2015} and \cite{Katanin2}.
At $T=0$ the spin stiffness is $\rho(0)=S|J|/8$.
This result for $\rho(0)$ for the pyrochlore ferromagnet 
should be contrasted to the result for the simple-cubic ferromagnet $\rho(0)=S|J|$.
The factor $1/8$ is easily understood by simple linear-spin-wave-theory arguments, see, e.g., Ref.~\cite{Nolting2009}. 
Indeed, in linear-spin-wave theory the stiffness is given by 
$\rho= (S/(2N))\sum_{i,j}J_{ij}({\bf{q}}\cdot{\bf{R}}_{ij})^2/\vert{\bf{q}}\vert^2$,
where the sum runs over all $N$ lattice sites,
however, $J_{ij}$ is nonzero only when ${\bf{R}}_{ij}$ connects the neighboring sites $i$ and $j$ on the lattice.
For the simple cubic lattice,
any site $i$ has six neighbors with ${\bf{R}}_{ij}=(\pm1,0,0)$, $(0,\pm1,0)$, $(0,0,\pm1)$ 
(i.e., the nearest-neighbor separation is $1$).
As a result, we get $\rho=S\vert J\vert$.
For the pyrochlore lattice,
we have to consider four different sites $i_1$, $i_2$, $i_3$, $i_4$ each of which has six neighbors,
and, most importantly, the nearest-neighbor separation is $1/\sqrt{8}$,  
see Sec.~\ref{sec2}.
That after all yields $\rho=S\vert J\vert/8$.

Above $T_c$, 
the small-wavevector excitation energies depend linearly on the wavevector with the excitation velocity $v=\sqrt{ J\Delta/24}$.
In the limit $T\to\infty$ we have $v=\vert J\vert\sqrt{S(S+1)/12}$ 
(i.e., $v=\vert J\vert/4$ for $S=1/2$).
For the $S=1/2$ simple-cubic Heisenberg ferromagnet 
the infinite-temperature value of excitation velocity is $v=\vert J\vert/\sqrt{2}$,
see Eqs.~(\ref{a03}), (\ref{a04}).
Again, 
the factor of $1/\sqrt{8}$ between the simple-cubic and the pyrochlore lattices 
is related to the difference in the nearest-neighbor separation.

In Fig.~\ref{fig03} we show the temperature dependences of the normalized spin stiffness and excitation velocity 
obtained from Eq.~(\ref{18}).
As it has been explained above,
the spin stiffness and the excitation velocity in the $S=1/2$ case 
are essentially smaller for the pyrochlore lattice than for the simple-cubic one 
due to the difference in the nearest-neighbor separation.
Moreover, both quantities decrease as $S$ increases.
In the inset in Fig.~\ref{fig03} we show $\rho/\rho(T=0)$ as a function of $T/\vert J\vert$.
This plot shows that the simple-cubic ferromagnet at $0 < T < T_c$ is more ``stiff'' in comparison to the pyrochlore one, 
i.e., there is an indication that the ferromagnetic phase in the pyrochlore ferromagnet is less stable against thermal fluctuations.

\subsection{Susceptibility, magnetization, critical temperature}
\label{sec4B}

\begin{figure}
\centering 
\includegraphics[clip=on,width=80mm,angle=0]{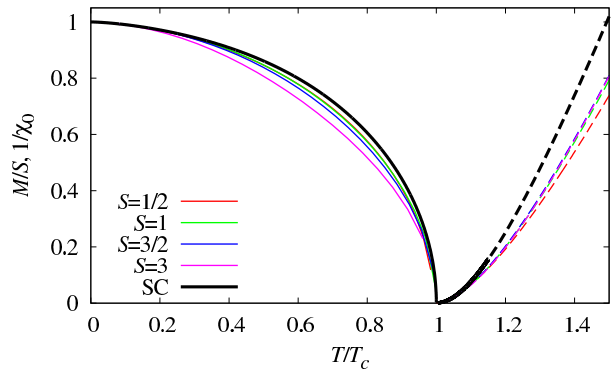}
\protect
\caption
{RGM data for the normalized magnetization $M/S$ of the ferromagnet on 
the simple-cubic lattice ($S=1/2$) (thick solid black line) 
and 
the pyrochlore lattice ($S=1/2,1,3/2,3$) (thin solid lines)
as a function of the normalized temperature $T/T_c$. 
Dashed curves correspond to the inverse uniform susceptibility $1/\chi_{\bf{0}}$ above $T_c$.
Note that the thin dashed curves  for $S>1/2$ almost coincide.}
\label{fig04} 
\end{figure}

\begin{figure}
\centering 
\includegraphics[clip=on,width=80mm,angle=0]{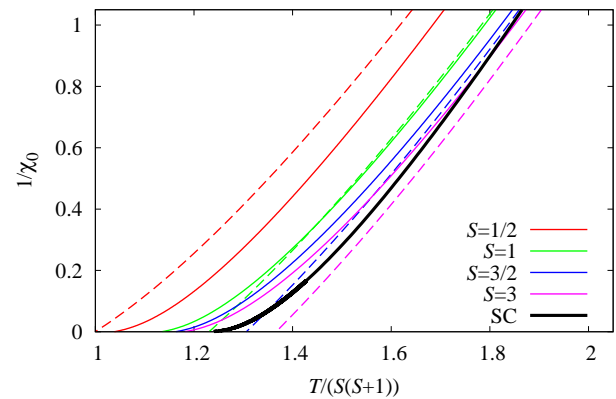}
\protect
\caption
{Inverse uniform susceptibility $1/\chi_{\bf{0}}$ of the ferromagnet on the pyrochlore lattice 
obtained by  the RGM (thin solid lines) and by the HTE  approach (Pad\'{e} [5,6] -- thin dashed lines) 
as a function of the normalized temperature $T/(S(S+1))$
for several spin quantum numbers $S$.
We also show the RGM results for the simple-cubic-lattice case with $S=1/2$ (thick black line).
Note that the  energy scale is set by $J=-1$.}
\label{fig05} 
\end{figure}

As it was already mentioned in Sec.~\ref{sec3A},
one straightforward outcome from the RGM equations 
is the uniform susceptibility $\chi_{\mathbf{0}}$ given in Eq.~(\ref{15}),
see Figs.~\ref{fig04} and \ref{fig05}.
Another straightforward outcome is the spontaneous magnetization $M$ (order parameter) related to the condensation term,
see Fig.~\ref{fig04}.
The temperature dependence of $\chi_{\mathbf{0}}$ or $M$ is used to determine the critical (Curie) temperature $T_c$.
Within the RGM,
$T_c$ follows from the equations $C_{\mathbf{0}} = 0$ and $\Delta=0$ 
(for the latter one, see Eq.~(\ref{16})).
In Fig.~\ref{fig04} we report the temperature dependences of the magnetization
as well as of the inverse uniform susceptibility.
According to these graphs,
for a fixed value of $T/T_c < 1$
the magnetization is larger for the $S=1/2$ simple-cubic ferromagnet than for the $S=1/2$ pyrochlore ferromagnet.
Comparing results for various spin quantum numbers $S$ we notice
that the magnetization decreases with further increasing of $S$ for the pyrochlore ferromagnet. 
Thus the $M(T/T_c)/S$ graphs for the pyrochlore case with large $S$ show a characteristic flattening. 
Note that this kind of flattening was also found to be a typical feature of disordered ferromagnets \cite{Handrich}.

\begin{figure}
\centering 
\includegraphics[clip=on,width=80mm,angle=0]{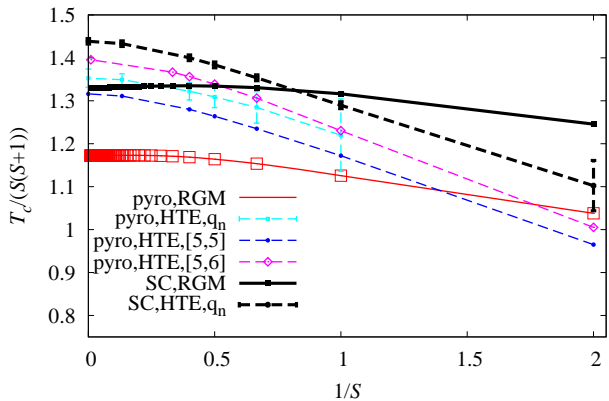}
\protect
\caption
{Normalized Curie temperatures $T_c/(S(S+1))$ of the ferromagnet on the simple-cubic lattice and the pyrochlore lattice 
within the RGM approach and the HTE  approach (up to the eleventh order) 
as a function of the inverse spin quantum number $1/S$. 
The HTE data labeled by ``pyro,HTE,[5,5]'' are taken from Ref.~\cite{Lohmann2014}.}
\label{fig06} 
\end{figure}

In Fig.~\ref{fig05} we compare the temperature dependences of $1/\chi_{\mathbf{0}}$ in some detail 
for different spin values $S$ obtained by RGM and HTE.
Although overall agreement of the two approaches is good,
there are noticeable differences in the  values of $T_c$ derived by the zeros of the inverse susceptibility, 
see also Fig.~\ref{fig06}.
Thus,
according to Fig.~\ref{fig06}
for the $S=1/2$ pyrochlore (simple-cubic) lattice
the RGM yields $T_c\approx 0.778$ ($T_c\approx 0.926$)
and  
the HTE (Pad\'{e} [5,5] and [5,6]) yields $T_c\approx 0.724 \ldots 0.754$ 
($T_c\approx 0.827$ \cite{Lohmann2014}).

As already discussed above, the  simple-cubic ferromagnet is  more ``stiff'' against thermal fluctuations, 
and, as a result, $T_c$ for the simple-cubic ferromagnet is obviously higher than that for the pyrochlore ferromagnet. 
Our results for the simple-cubic case may be compared to the quantum Monte Carlo result $T_c=0.839(1)$ \cite{Wessel2010}.
We have also performed quantum Monte Carlo simulations for the $S=1/2$ pyrochlore ferromagnet 
using ALPS package (looper algorithm) \cite{ALPS} 
and found $T_c\approx 0.718$. 

A similar comparison can be performed in the other limiting case $S\to\infty$.
In this case 
for the pyrochlore lattice
the RGM yields $T_c/(S(S+1)) \approx 1.172$,
the HTE (Pad\'{e} [5,5] and [5,6]) yields $T_c/(S(S+1)) \approx 1.316 \ldots 1.396$,
whereas
the classical Monte Carlo simulations yield $T_c/(S(S+1))\approx 1.317$ \cite{Soldatov2017}.
For the simple-cubic lattice we have 
$T_c/(S(S+1)) \approx 1.317$ (RGM),
$T_c/(S(S+1)) \approx 1.438$ (HTE) \cite{Lohmann2014},
$T_c/(S(S+1)) \approx 1.443$ (classical Monte Carlo \cite{Peczak1991,Soldatov2017}),
respectively.
Although there is some variance in the values of $T_c$ obtained by different methods,
all results indicate that the Curie temperature of the pyrochlore Heisenberg ferromagnet 
is about 85\% ($S=1/2$) or about 90\% ($S\to\infty$) of the Curie temperature of the simple-cubic Heisenberg ferromagnet.
For convenience, we have collected these data for $T_c$ in Tables~\ref{table1} and \ref{table2}. 

\begin{table}
\centering
\caption{Critical temperature $T_c$  for the  quantum ($S=1/2$) pyrochlore and simple-cubic Heisenberg ferromagnets 
($\vert J\vert=1$).}
\label{table1}
\begin{tabular}{c||c|c}
Method            & Pyrochlore lattice & Simple-cubic lattice\\
\hline
RGM               & 0.778              & 0.926\\
\hline
HTE (Pad\'{e})    & 0.724\ldots 0.754  & 0.827\\
\hline
QMC               & 0.718              & 0.839(1)\\
\end{tabular}
\end{table}

\begin{table}
\centering
\caption{Critical temperature $T_c/(S(S+1))$ for the classical ($S\to\infty$) pyrochlore and simple-cubic Heisenberg ferromagnets ($\vert J\vert=1$).
The corresponding results for the $S=1/2$ (quantum) case are given in parentheses.}
\label{table2}
\begin{tabular}{c||c|c}
Method             & Pyrochlore lattice  & Simple-cubic lattice\\
\hline
RGM                & 1.172               & 1.317\\
                   & (1.037)             & (1.235)\\
\hline
HTE (Pad\'{e})     & 1.316\ldots 1.396   & 1.438\\
                   & (0.965\ldots 1.005) & (1.103)\\
\hline
CMC                & 1.317               & 1.443\\
(QMC)              & (0.957)             & (1.119)\\
\end{tabular}
\end{table}

We mention further, 
that the Curie temperature $T_c$ of the $S=1/2$ pyrochlore ferromagnet was determined previously to $T_c=0$ 
using a phenomenological renormalization group method \cite{Garcia-Adeva2014}.
This result  is certainly an artefact of the applied approach.

An important quantity which can be obtained from the ${\bf{q}}$-dependent susceptibility
\begin{eqnarray}
\label{19}
\chi_{\mathbf{q}}
=
\underset{\omega\rightarrow 0}{\textrm{lim}}\frac{1}{8}\sum_{\alpha,\beta}\chi_{\mathbf{q}\alpha\beta}^{+-}(\omega) 
\end{eqnarray}
is the correlation length $\xi_{\mathbf{Q}}$.
By expanding the susceptibility $\chi_\mathbf{q}$ (\ref{19}) around the magnetic order
wavevector ${\mathbf{Q}=\mathbf{0}}$
we get 
$\chi_{\mathbf{Q}+\Delta{\bf{q}}}\approx\chi_\mathbf{Q}/(1+\xi_\mathbf{Q}^2(\Delta\mathbf{q})^2)$, 
see, e.g., Refs.~\cite{Haertel2010,Junger2005,Mueller2015,Schmalfus2004}.
We find $\xi_{\mathbf{0}}=\sqrt{|J|\alpha_{100}\chi_{\mathbf{0}}/8}$ for the pyrochlore ferromagnet.
The RGM approach for the simple-cubic ferromagnet yields the value
$\xi_{\mathbf{0}} =\sqrt{|J|\alpha_{100}\chi_{\mathbf{0}}}$.
Clearly, because of these relations between $\xi_{\mathbf{0}}$ and $\chi_{\mathbf{0}}$,  
the qualitative behavior of the correlation length as a function of temperature can be estimated from Figs.~\ref{fig04} and \ref{fig05}.

Summarizing the discussion of the temperature dependences of the magnetization and of the susceptibility 
as well as the results of the critical temperature,
we again may conclude that the ferromagnetic phase in the pyrochlore ferromagnet is stronger affected by temperature fluctuations,
which can be related to the frustrated geometry of the pyrochlore lattice, 
see the general discussion of this issue in the introduction.

\subsection{Spin-spin correlation functions, specific heat, and structure factor}
\label{sec4C}

\begin{figure}
\centering 
\includegraphics[clip=on,width=80mm,angle=0]{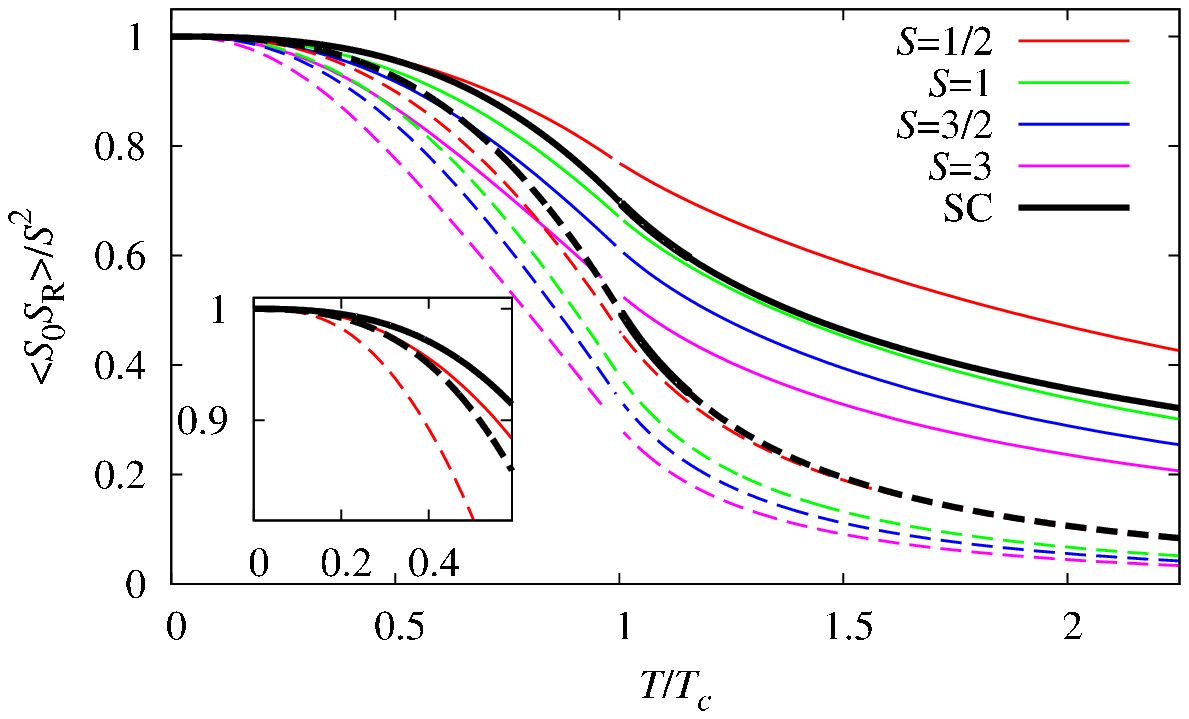}
\protect
\caption
{Main panel: 
normalized correlation functions $\langle \hat{{\bf{S}}}_{\mathbf{0}}\cdot\hat{{\bf{S}}}_{\mathbf{R}}\rangle/S^2$ 
(nearest neighbors -- solid; next-nearest neighbors -- dashed)
as a function of the normalized temperature $T/T_c$ 
for the spin-$S$ pyrochlore ferromagnet for several spin quantum numbers $S$ (thin lines).
We also show the results for the $S=1/2$ simple-cubic ferromagnet (thick lines).
Inset: 
$\langle \hat{{\bf{S}}}_{\mathbf{0}}\cdot\hat{{\bf{S}}}_{\mathbf{R}}\rangle/S^2$ versus $T/\vert J\vert$ for the $S=1/2$ case.}
\label{fig07} 
\end{figure}

The RGM approach yields straightforwardly spin-spin correlation functions, 
see Eqs.~(\ref{12}) and (\ref{13}).
In Fig.~\ref{fig07} we show the temperature dependences of the normalized correlation functions, 
$\langle \hat{{\bf S}}_0\cdot \hat{{\bf S}}_{\bf R}\rangle/S^2$, 
for nearest-neighbor and next-nearest-neighbor separations for $S=1/2,1,3/2$, and $3$. 
As increasing of $S$ 
the decrease of $\langle \hat{{\bf S}}_0\cdot \hat{{\bf S}}_{\bf R}\rangle/S^2$ with growing temperature becomes faster. 
As expected, 
the decay of the next-nearest-neighbor correlations is more rapid, 
but above $T_c$ the pronounced short-range order is obvious.  
Furthermore,
in the inset of Fig.~\ref{fig07} we compare results for the $S=1/2$ simple-cubic and pyrochlore ferromagnets 
showing the dependence of nearest-neighbor (solid) and next-nearest-neighbor (dashed) correlation functions 
as a function of $T/\vert J\vert$.
Again, 
with increasing temperature, the correlations for the pyrochlore lattice vanish more rapidly than for the simple-cubic lattice.
Since the nearest-neighbor correlation function is proportional to the internal energy of the spin model (\ref{01})
the temperature profiles reported in Fig.~\ref{fig07} represent also the temperature dependence of the internal energy.

\begin{figure}
\centering 
\includegraphics[clip=on,width=80mm,angle=0]{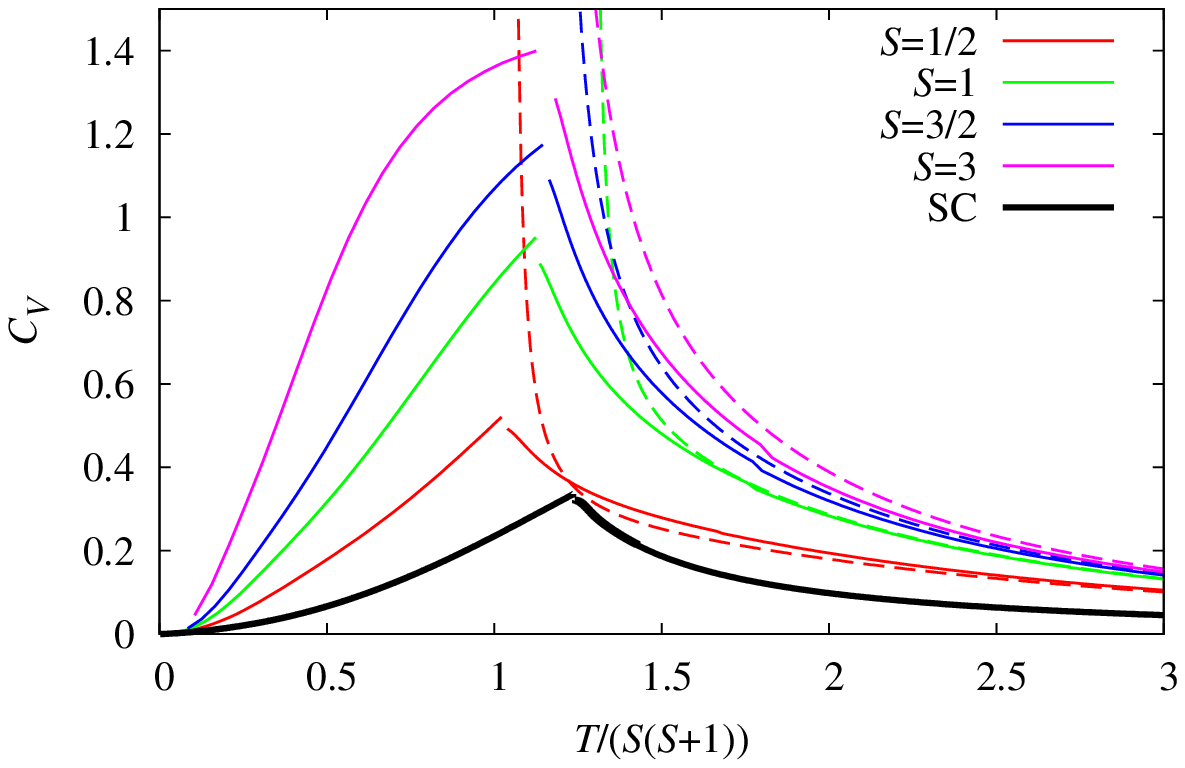}
\protect
\caption
{Specific heat of the ferromagnet on the pyrochlore lattice 
within the RGM (thin solid lines) and the HTE  approach (Pad\'{e} [5,6] -- thin dashed lines) 
as a function of the normalized temperature $T/(S(S+1))$
for several values of the spin quantum numbers $S$.
We also show the RGM results for the $S=1/2$ simple-cubic ferromagnet (thick solid line).
}
\label{fig08} 
\end{figure}

Next we present the temperature dependence of the specific heat $C_V(T)$, see Fig.~\ref{fig08}. 
It shows the typical cusp at $T_c$.
We also show the HTE results for the high-temperature part of $C_V(T)$:
they begin to rise sharply as the temperature approaches $T_c$ from above, thus, indicating the phase transition.
In the high-temperature region the HTE and the RGM results coincide.

\begin{figure}
\centering 
\includegraphics[clip=on,width=42mm,angle=0]{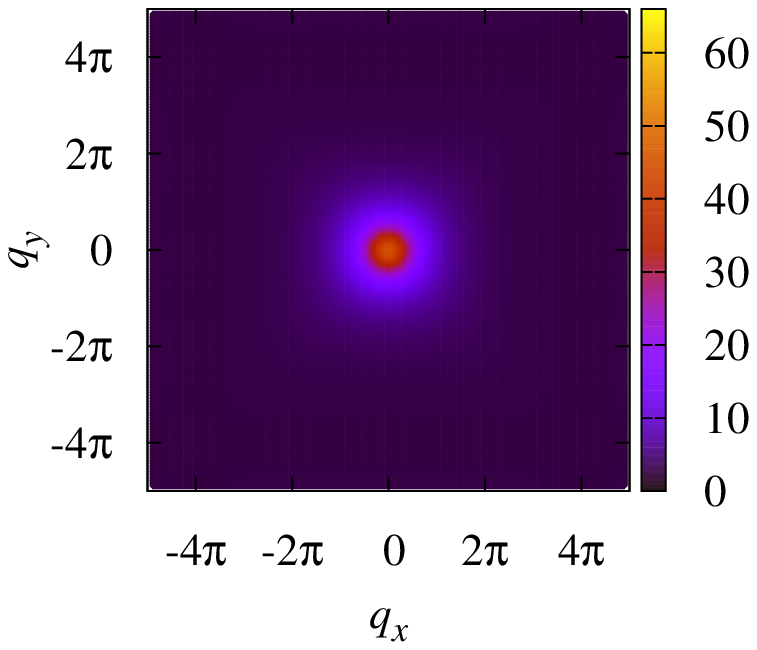}
\includegraphics[clip=on,width=42mm,angle=0]{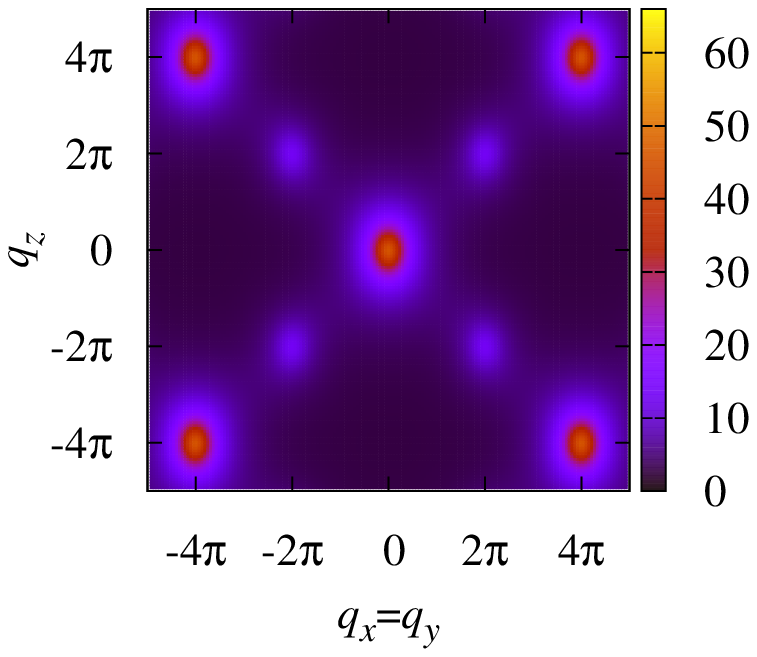}\\
\includegraphics[clip=on,width=42mm,angle=0]{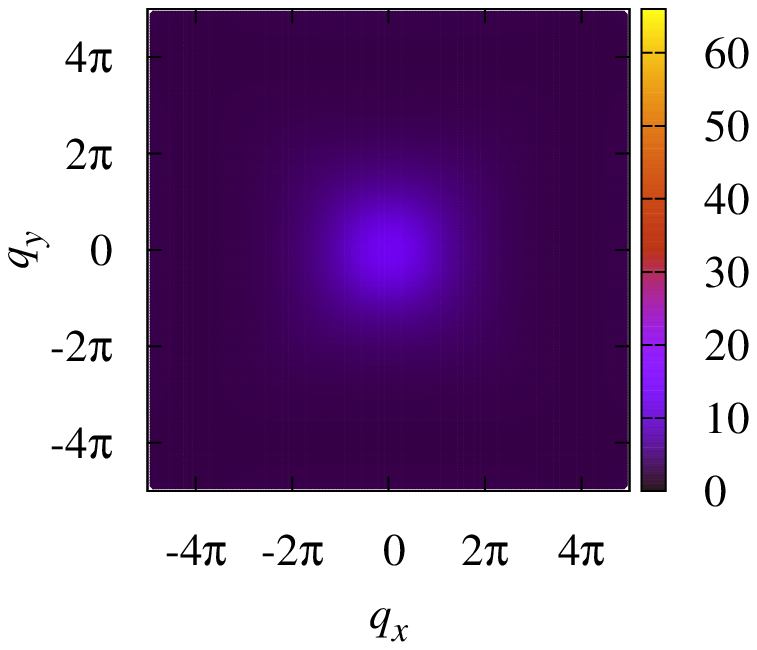}
\includegraphics[clip=on,width=42mm,angle=0]{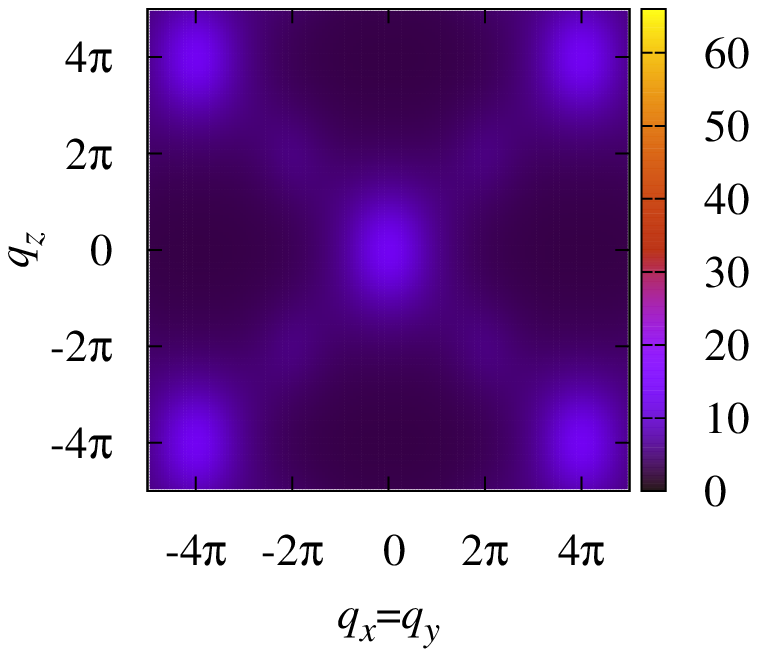}\\
\vspace{3mm}
\includegraphics[clip=on,width=42mm,angle=0]{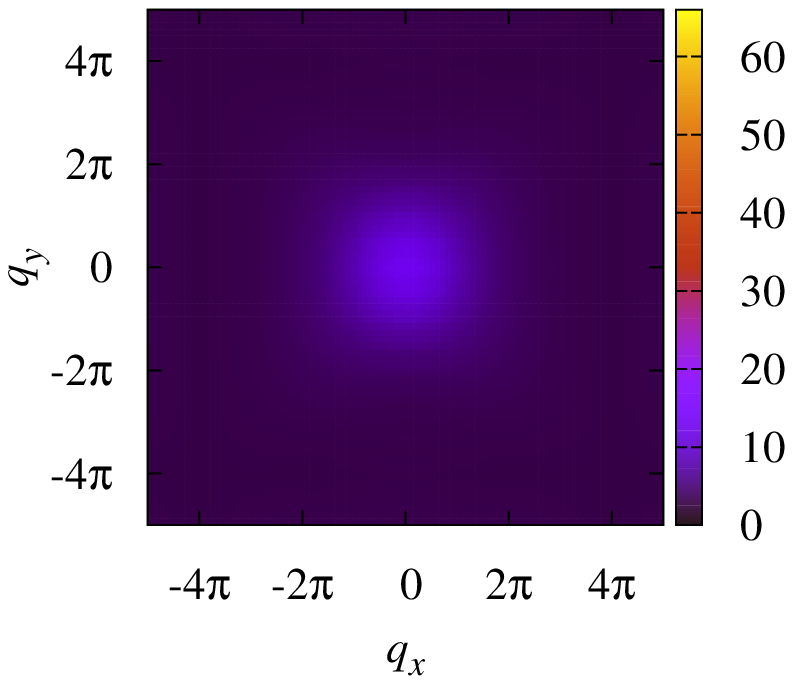}
\includegraphics[clip=on,width=42mm,angle=0]{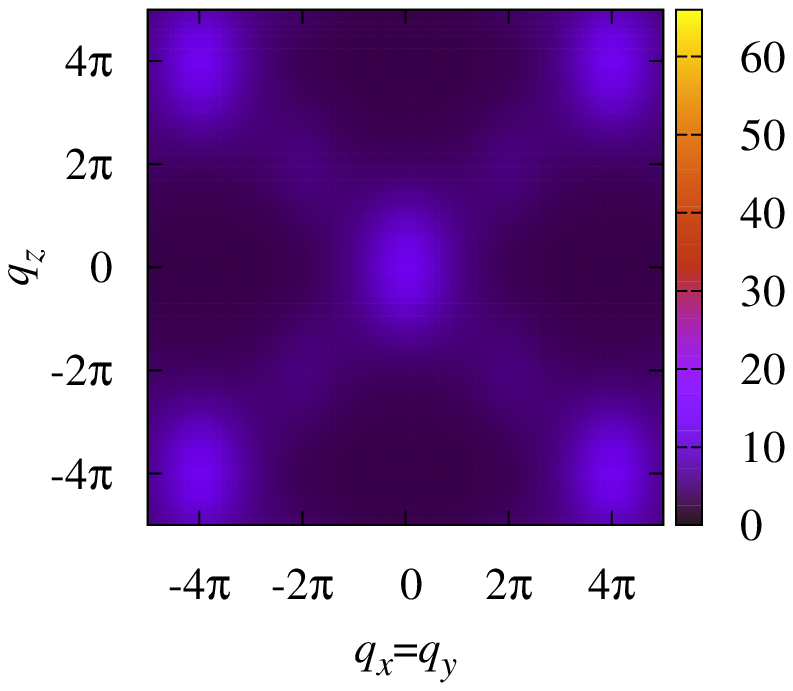}
\protect
\caption
{Two top rows:
magnetic structure factor $S_{\mathbf{q}}/(S(S+1))$ of the $S=1/2$ ferromagnet on the  pyrochlore lattice 
within the RGM approach at $T=1.3 T_c$ (upper row) and $T=2 T_c$ (middle row) 
in the Bragg plane $q_z=0$ (left panels) and in the Bragg plane $q_x=q_y$ (right panels).
Bottom row:
magnetic structure factor $S_{\mathbf{q}}/(S(S+1))$ of the $S=1/2$ ferromagnet on the pyrochlore lattice 
within the HTE approach (ninth order) at $T=2 T_c$ 
in the Bragg plane $q_z=0$ (left panel) and in the Bragg plane $q_x=q_y$ (right panel).}
\label{fig09} 
\end{figure}

\begin{figure}
\centering 
\includegraphics[clip=on,width=80mm,angle=0]{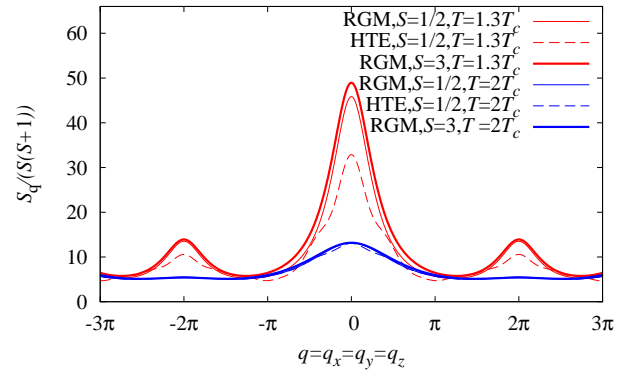}
\protect
\caption
{Magnetic structure factor $S_{\mathbf{q}}/(S(S+1))$ of the pyrochlore ferromagnet along the line $q_x=q_y=q_z$
for two temperatures: $1.3T_c$ (red) and $2T_c$ (blue).
RGM results are shown by solid lines, whereas HTE results are shown by dashed lines.
Thin lines correspond to the $S=1/2$ case, 
thick lines correspond to $S=3$ case.}
\label{fig10} 
\end{figure}

The correlation functions provide the access to the (static) magnetic structure factor (\ref{17})
which is related to an experimentally accessible quantity, 
the total magnetic neutron cross section ${\rm{d}}\sigma/{\rm{d}}\Omega$.
We present a contour plot of the structure factor in several planes of the ${\bf{q}}$-space,
namely,
$q_z=0$ (left panels of Fig.~\ref{fig09})
and
$q_x=q_y$ (right panels of Fig.~\ref{fig09}). 
In Fig.~\ref{fig09} we also compare the RGM and HTE predictions (above $T_c$) shown in the middle and bottom rows, respectively.
Clearly, the results of both approaches are in good agreement.
To get a more quantitative profile of the structure factor 
we present the dependence of $S_{(q,q,q)}$ on $q$ for two values of $T$ and $S=1/2$ and $S=3$ in Fig.~\ref{fig10}. 
This $q$-line corresponds to a diagonal line in the right panels of Fig.~\ref{fig09}.

As expected for ferromagnets, the magnetic structure factor has rather simple features:
it exhibits a pronounced  maximum around the $\Gamma$ point ${\bf{q}}=(0,0,0)$.
However, along the path $q_x=q_y=q_z$  
a second maximum appears, see Fig.~\ref{fig10}
and the right panels of  Fig.~\ref{fig09}.
For a better  understanding of  
the shape of $S_{{\bf{q}}}$ 
we return to the definition of the magnetic structure factor
and rewrite  Eq.~(\ref{17}) as a sequence of contributions 
coming from on-site correlations, nearest-neighbor correlations, next-nearest-neighbor correlations, third-neighbor correlations etc., 
i.e.,
\begin{eqnarray}
\label{20}
S_{{\bf{q}}}
=
S(S+1)
+\langle \hat{\bf{S}}_{{\bf{0}}} \cdot \hat{\bf{S}}_{{\bf{1}}} \rangle S^{(1)}_{{\bf{q}}}
+\langle \hat{\bf{S}}_{{\bf{0}}} \cdot \hat{\bf{S}}_{{\bf{2}}} \rangle S^{(2)}_{{\bf{q}}}
+\ldots,
\nonumber\\
S^{(1)}_{{\bf{q}}}=\frac{1}{4}\sum_{\alpha}\sum_{j^\prime}\cos\left({\bf{q}}\cdot\left({\bf{R}}_{m\alpha}-{\bf{R}}_{j^\prime}\right) \right),
\quad
\nonumber\\
S^{(2)}_{{\bf{q}}}=\frac{1}{4}\sum_{\alpha}\sum_{j^{\prime\prime}}\cos\left({\bf{q}}\cdot\left({\bf{R}}_{m\alpha}-{\bf{R}}_{j^{\prime\prime}}\right) \right),
\quad
\end{eqnarray}
where
$\langle \hat{\bf{S}}_{{\bf{0}}} \cdot \hat{\bf{S}}_{{\bf{1}}} \rangle$ is the nearest-neighbor correlation function,
$\langle \hat{\bf{S}}_{{\bf{0}}} \cdot \hat{\bf{S}}_{{\bf{2}}} \rangle$ is the next-nearest-neighbor correlation function
etc.,
and the sum over $j^\prime$ runs over nearest neighbors of the site $i=m\alpha$,
the sum over $j^{\prime\prime}$ runs over next-nearest neighbors of the site $i=m\alpha$ etc.
Considering the path  along ${\bf{q}}=(q,q,q)$,
one can easily explain the dependence of the magnetic structure factor on $q$ shown in Fig.~\ref{fig10}.
Really, since 
$S^{(1)}_{(0,0,0)}=6$,  
$S^{(1)}_{(\pi,\pi,\pi)}=S^{(1)}_{(2\pi,2\pi,2\pi)}=S^{(1)}_{(3\pi,3\pi,3\pi)}=0$, 
$S^{(1)}_{(4\pi,4\pi,4\pi)}=6$,  
the nearest-neighbor correlations 
contribute to $S_{{\bf{q}}}$ at ${\bf{q}}=(0,0,0)$ and ${\bf{q}}=(4\pi,4\pi,4\pi)$,
but do not contribute at ${\bf{q}}=(2\pi,2\pi,2\pi)$.
Furthermore, 
$S_{{\bf{q}}}$ at ${\bf{q}}=(2\pi,2\pi,2\pi)$ is conditioned first of all by much weaker third-neighbor (next-next-nearest-neighbor) correlations, 
since 
$S^{(2)}_{(2\pi,2\pi,2\pi)}=0$, 
but
$S^{(3)}_{(2\pi,2\pi,2\pi)}=6$.
As a result, the dependence $S_{(q,q,q)}$ on $q$ shows a high maximum at $q=0$ (and $q=4\pi$) and a lower one at $q=2\pi$.
Naturally, the heights of the maxima at $q=0$ and $q=2\pi$ increase as the temperature decreases.

We end up with few further comments on the $q$-dependence shown in Fig.~\ref{fig10}.
Comparing thin solid ($S=1/2$) and thick solid ($S=3$) lines 
we conclude that the peaks of $S_{{\bf{q}}}/(S(S+1))$, especially at $q=0$, become higher as $S$ increases.
Comparing thin solid (RGM) and thin dashed (HTE) lines at two temperatures,
$T=1.3T_c$ (red) and $T=2T_c$ (blue),
we conclude that in general RGM and HTE results are in a reasonable agreement 
and the agreement becomes better at higher temperatures.

\subsection{Dynamic structure factor}
\label{sec4D}

The RGM results given in Eq.~(\ref{11}) allow to determine the dynamic structure factor 
using the fluctuation-dissipation theorem,
see the end of Sec.~\ref{sec3A}.
After some standard manipulations we arrive at
\begin{eqnarray}
\label{21}
S_{{\bf{q}}}^{zz}(\omega)
=
\frac{\pi}{1-e^{-\frac{\omega}{T}}}\sum_{\alpha,\beta}\sum_{\gamma}\frac{m_{\gamma{\bf{q}}}}{8\omega_{\gamma{\bf{q}}}}
\nonumber\\
\times
\big(
\delta(\omega-\omega_{\gamma{\bf{q}}})
-
\delta(\omega+\omega_{\gamma{\bf{q}}})
\big)
\langle \alpha\vert\gamma{\bf{q}}\rangle \langle \gamma{\bf{q}}\vert\beta\rangle.
\end{eqnarray}
This quantity is related to neutron inelastic scattering data accessible in experiments.
We also note that integrating $S_{{\bf{q}}}^{zz}(\omega)$ (\ref{21}) over all $\omega$ 
we get the static structure factor:
\begin{eqnarray}
\label{22}
\int_{-\infty}^{\infty}{\rm{d}}\omega S_{{\bf{q}}}^{zz}(\omega)
=
2\pi S_{{\bf{q}}}^{zz}
=
2\pi\frac{1}{3}S_{{\bf{q}}}.
\end{eqnarray}
In our numerical calculation we replace the $\delta$-functions in Eq.~(\ref{21}) by the Lorentzian function, 
i.e., $\delta(x)\to (1/\pi)(\epsilon/(x^2+\epsilon^2))$, 
where a ``damping'' parameter $\epsilon$ is chosen as $\epsilon = 0.001\ldots 0.5$.
(Note that there is no intrinsic damping in the RGM approach.)

\begin{figure}
\centering 
\includegraphics[clip=on,width=80mm,angle=0]{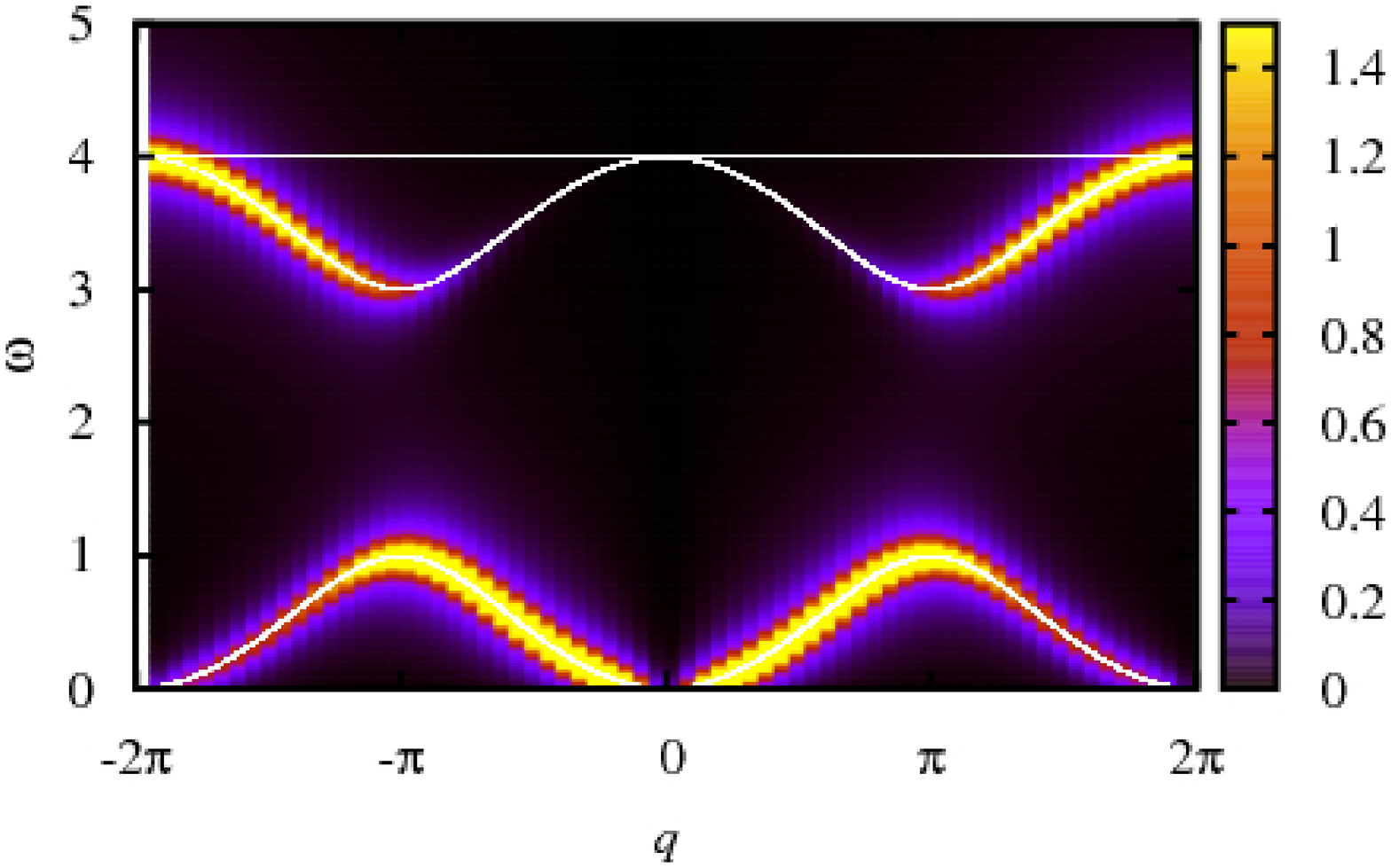}
\includegraphics[clip=on,width=80mm,angle=0]{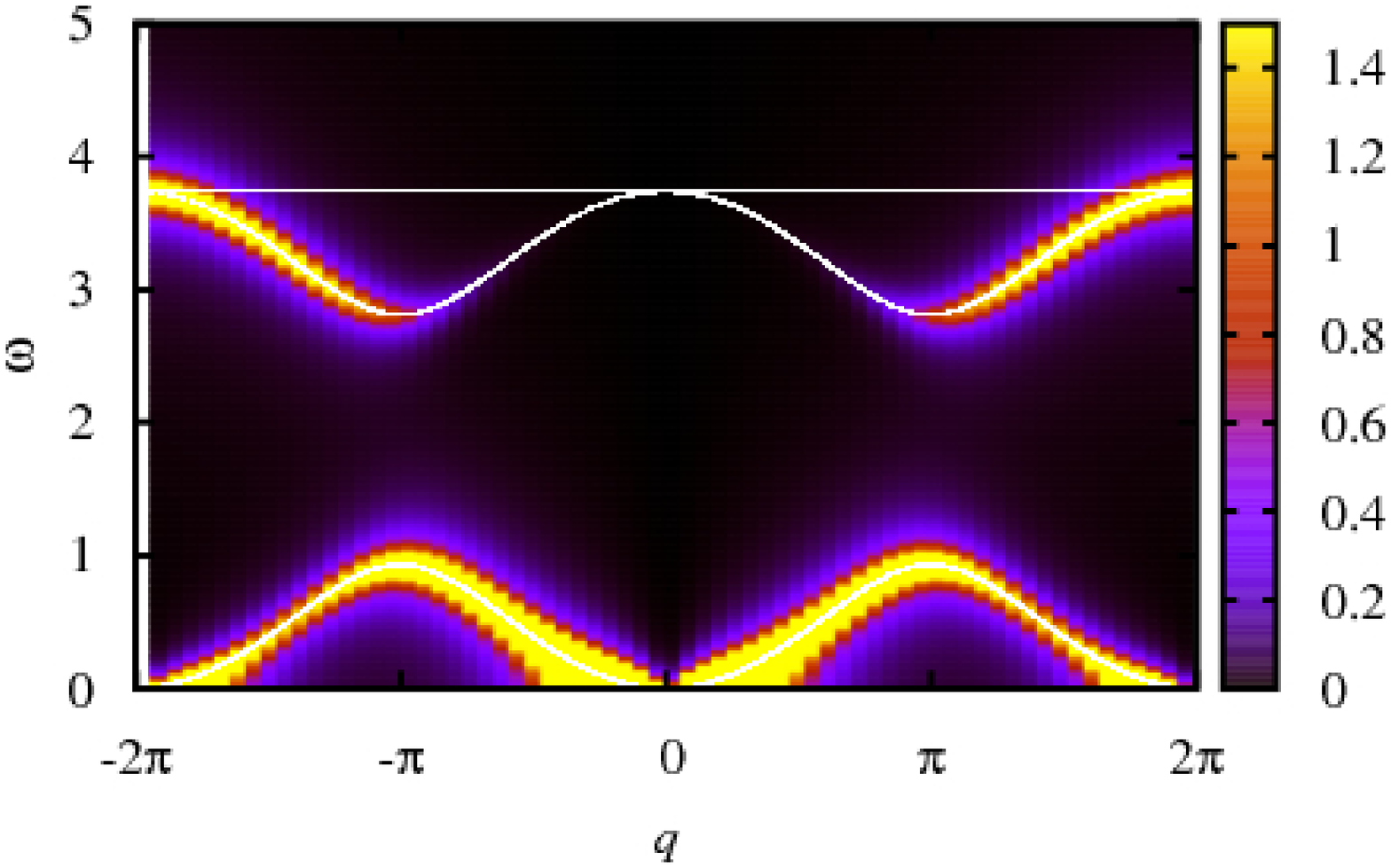}
\protect
\caption
{Dynamic structure factor $S^{zz}_{\mathbf{q}}(\omega)$ of the $S=1/2$ pyrochlore ferromagnet along the line $q_x=q_y=q_z$ 
for $T=0.0425$ (top) and $T=0.425$ (bottom). 
We set $\epsilon=0.1$.
The white lines correspond to the excitation energies $\omega_{\gamma{\mathbf{q}}}$ (\ref{09}).}
\label{fig11} 
\end{figure}

In Fig.~\ref{fig11} we show  $S^{zz}_{\mathbf{q}}(\omega)$, Eq.~(\ref{21}), 
in the wavevector $q=q_x=q_y=q_z$ -- frequency $\omega$ plane,
(cf. right panels in Fig.~\ref{fig09} and Fig.~\ref{fig10})
for the $S=1/2$ case at the temperatures $T=0.0425$ (top) and $T=0.425$ (bottom).
The temperature value $0.0425\vert J\vert$ is related to experimental data of Ref.~\cite{Mena2014}:
if $J=8.22$~meV then $0.0425\vert J\vert$ corresponds to 4~K
(and $0.425\vert J\vert$ corresponds to 40~K).
We also plot by white lines the excitation energies $\omega_{\gamma{\bf{q}}}$ (\ref{09}) along the line $q=q_x=q_y=q_z$
(cf. Fig.~\ref{fig02}).
Evidently,
$S^{zz}_{\bf{q}}(\omega)$ is concentrated along the excitation energy lines $\omega_{\gamma{\bf{q}}}$.
However, its weight is distributed nonuniformly 
and is mostly concentrated along the acoustic branch $\omega_{4{\bf{q}}}$ and the branch $\omega_{3{\bf{q}}}$,
whereas high-energy flat-band branches $\omega_{1{\bf{q}}}$ and $\omega_{2{\bf{q}}}$ are not visible.

\begin{figure}
\centering 
\includegraphics[clip=on,width=80mm,angle=0]{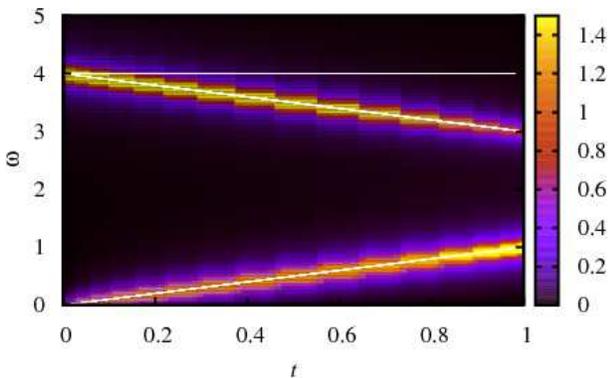}
\protect
\caption
{Dynamic structure factor $S^{zz}_{\mathbf{q}}(\omega)$ of the $S=1/2$ pyrochlore ferromagnet 
as a function of the reduced momentum $t=2-D_{\mathbf{q}}$ with $\mathbf{q}=(q,q,q)$ 
for $T=0.0425$. 
We set $\epsilon=0.1$.
The white lines correspond to the excitation energies $\omega_{\gamma{\mathbf{q}}}$ (\ref{09}).}
\label{fig12} 
\end{figure}

To get a closer relation to the experimental paper \cite{Mena2014} on the pyrochlore ferromagnet Lu$_2$V$_2$O$_7$,
we show in Fig.~\ref{fig12}  the dynamic structure factor 
as a function of the reduced momentum $t=2-D_{{\bf{q}}}$,
see Eq.~(\ref{08}),
along the path ${\bf{q}}=(q,q,q)$, which  corresponds to a diagonal line in the right panels of Fig.~\ref{fig09}. 
Then the reduced momentum $t$ varies between 0 and 1.
Note that Fig.~\ref{fig12} resembles Fig.~2 of Ref.~\cite{Mena2014}. 
However, in the experimental paper \cite{Mena2014} 
an average over many ${\bf q}$-points lying within a sphere around the $\Gamma$-point of a given radius is performed.

By comparing experimental neutron inelastic scattering data with theoretical predictions for $S_{{\bf{q}}}^{zz}(\omega)$ 
the parameters of the Hamiltonian 
(i.e., the value of the nearest-neighbor exchange coupling) 
can be determined for a certain magnetic compound.
In the case at hand,
it is natural to consider the highest experimentally observed energies 
(around $\omega\approx 4\vert J\vert$) 
to get the value of $J$.
Comparing the results for two different temperatures, 
e.g., $T=0.0425$ and $T=0.425$, 
see Fig.~\ref{fig11},
one can estimate temperature effects which influence the determination of $J$.

\section{Summary}
\label{sec5} 

To summarize,
we have presented a comprehensive study of finite-temperature static and dynamic properties 
of the spin-$S$ pyrochlore Heisenberg ferromagnet for arbitrary $S\ge 1/2$.
In particular,
we focus on the excitation spectra, the susceptibility, the magnetization, the specific heat 
as well as the static and dynamic structure factors.
The reported results were obtained within the frames of two universal
approaches, 
the rotation-invariant Green's function method and the high-temperature expansion.
To demonstrate the effect of geometric frustration on the finite-temperature properties of the pyrochlore ferromagnet, 
we compare the pyrochlore and the simple-cubic ferromagnets.
Overall,
the difference between thermodynamics of the pyrochlore ferromagnet and the simple-cubic ferromagnet is noticeable, 
although it is not tremendous.

Our results may be used for understanding experimental data for Heisenberg pyrochlore ferromagnets at finite temperatures,
see Refs.~\cite{Zhou2008,Onose2010,Mena2014,Yasui2003,Menyuk1966,Wojtowicz1969,Yaresko2008,Tymoshenko2017}.
Con\-cer\-ning ferromagnetic pyrochlore compounds,
we have mentioned already in Sec.~\ref{sec1} that for the $S=1/2$ Heisenberg ferromagnet on the pyrochlore lattice Lu$_2$V$_2$O$_7$ \cite{Zhou2008,Onose2010,Mena2014}
the critical temperature is $T_c \approx 0.73\vert J\vert$. 
For another compound, Yb$_2$Ti$_2$O$_7$, 
with a much lower critical temperature of about $0.24$~K \cite{Yasui2003},
one finds $T_c \approx 0.68 \vert J\vert$.
The ratio $T_c/\vert J\vert \approx 0.7$ agrees well with our theoretical findings.
Note, 
however, 
that the low symmetry of the pyrochlore lattice allows for a (typically weak) Dzyaloshinskii-Moriya interaction. 
Furthermore, 
inelastic neutron scattering data for $\omega$ of the order of $J$ reveal excitations of the spin system.
Comparing experimental data and theoretical predictions allows one to determine the model parameters.
In contrast to linear-spin-wave-theory calculations of the excitation energy dispersion,
the RGM findings for the dynamic structure factor are not limited to the low-temperature limit. 

\section*{Acknowledgments}

The authors thank M.~E.~Zhitomirsky for discussions and correspondence.
The present study was supported by the Deutsche Forschungsgemeinschaft (project RI615/21-2).
O.~D. acknowledges the kind hospitality of the University of Magdeburg in October-December of 2016 and April-May of 2017.
The work of O.~D. was partially supported by Project FF-30F (No.~0116U001539) from the Ministry of Education and Science of Ukraine.
O.~D. would like to thank the Abdus Salam International Centre for Theoretical Physics (Trieste, Italy) 
for partial support of these studies through the Senior Associate award.

\section*{Appendix A: RGM results for the $S=1/2$ simple-cubic Heisenberg model}
\renewcommand{\theequation}{A\arabic{equation}}
\setcounter{equation}{0}

In this appendix,
we present an analogue of Eqs.~(\ref{04}), (\ref{05}), and (\ref{06}) 
for the $S=1/2$ simple-cubic Heisenberg model 
(see also Refs.~\cite{Kawabe1973,preprint}).
For this case we have:
\begin{eqnarray}
\label{a01}
(\omega^2 - F_{\mathbf{q}})\chi^{{+-}}_{\mathbf{q}}(\omega) = -M_{\mathbf{q}},
\end{eqnarray}
where
\begin{eqnarray}
\label{a02}
\frac{M_{\mathbf{q}}}{J} = -12c_{100}(1-\gamma_{\bf{q}})
\end{eqnarray}
and
\begin{eqnarray}
\label{a03}
\frac{F_{\mathbf{q}}}{J^{2}}  
=  
3\left(1-\gamma_{\bf{q}}\right)
\left(1+10\tilde{\alpha}_{100} + 8\tilde{\alpha}_{110}+ 2\tilde{\alpha}_{200}
\right.
\nonumber\\
\left.
-12\tilde{\alpha}_{100}\left(1+\gamma_{\bf{q}}\right)\right)
\end{eqnarray}
with
\begin{eqnarray}
\label{a04}
\gamma_{\bf{q}}=\frac{1}{3}\left(\cos q_x + \cos q_y + \cos q_z\right).
\end{eqnarray}
Equation~(\ref{a01}) immediately yields 
$\chi^{{+-}}_{\mathbf{q}}(\omega)=-M_{\mathbf{q}}/(\omega^2 -\omega_{\bf{q}}^2)$
with $\omega_{\bf{q}}^2=F_{\mathbf{q}}$
(cf. Eq.~(\ref{11})).
Further calculations go parallel with the corresponding ones for the pyrochlore case.

\section*{Appendix B: Common eigenvectors $\vert\gamma{\bf{q}}\rangle$ of the momentum and frequency matrices}
\renewcommand{\theequation}{B\arabic{equation}}
\setcounter{equation}{0}

In this appendix,
we present the common eigenvectors $\vert\gamma{\bf{q}}\rangle$ 
of the momentum matrix $M_{{\bf{q}}}$ (\ref{05}) and the frequency matrix $F_{{\bf{q}}}$ (\ref{06}).
They are as follows:
\begin{widetext}
\begin{eqnarray}
\label{b01}
\vert 1{\bf{q}}\rangle
=
\left(
\begin{array}{c}
-\frac{\sin\frac{q_{x}-q_{z}}{4}}{\sin\frac{q_{x}+q_{y}}{4}}\\
-\frac{\sin\frac{q_{y}+q_{z}}{4}}{\sin\frac{q_{x}+q_{y}}{4}}\\
0\\
1
\end{array}
\right),
\end{eqnarray}
\begin{eqnarray}
\label{b02}
\vert 2{\bf{q}}\rangle
=
\left(
\begin{array}{c}
-\frac{\sin\frac{q_{y}-q_{z}}{4}}{\sin\frac{q_{x}+q_{y}}{4}}\\
-\frac{\sin\frac{q_{x}+q_{z}}{4}}{\sin\frac{q_{x}+q_{y}}{4}}\\
1\\
0
\end{array}
\right),
\end{eqnarray}
\begin{eqnarray}
\label{b03}
\vert 3{\bf{q}}\rangle
=
\left(
\begin{array}{c}
\frac{2D_{\mathbf{q}}\sin\frac{q_{y}+q_{z}}{4}+\sin\frac{2q_{x}+q_{y}-q_{z}}{4}+\sin\frac{2q_{x}-q_{y}+q_{z}}{4}}{\sin\frac{q_{x}-q_{y}}{2}+\sin\frac{q_{x}-q_{z}}{2}-\sin\frac{q_{y}+q_{z}}{2}}\\
-\frac{2\left(D_{\mathbf{q}}\sin\frac{q_{x}-q_{z}}{4}+\sin\frac{q_{y}}{2}\cos\frac{q_{x}+q_{z}}{4}\right)}{\sin\frac{q_{x}-q_{y}}{2}+\sin\frac{q_{x}-q_{z}}{2}-\sin\frac{q_{y}+q_{z}}{2}}\\
-\frac{2\left(D_{\mathbf{q}}\sin\frac{q_{x}-q_{y}}{4}+\sin\frac{q_{z}}{2}\cos\frac{q_{x}+q_{y}}{4}\right)}{\sin\frac{q_{x}-q_{y}}{2}+\sin\frac{q_{x}-q_{z}}{2}-\sin\frac{q_{y}+q_{z}}{2}}\\
1
\end{array}
\right),
\end{eqnarray}
\begin{eqnarray}
\label{b04}
\vert 4{\bf{q}}\rangle
=
\left(
\begin{array}{c}
\frac{-2D_{\mathbf{q}}\sin\frac{q_{y}+q_{z}}{4}+\sin\frac{2q_{x}+q_{y}-q_{z}}{4}+\sin\frac{2q_{x}-q_{y}+q_{z}}{4}}{\sin\frac{q_{x}-q_{y}}{2}+\sin\frac{q_{x}-q_{z}}{2}-\sin\frac{q_{y}+q_{z}}{2}}\\
\frac{4\cos\frac{q_{z}}{2}\cos\frac{q_{x}-q_{y}}{4}\left(D_{\mathbf{q}}+\cos\frac{q_{x}+q_{y}}{2}+3\right)+4\cos\frac{q_{x}+q_{y}}{4}\left(3D_{\mathbf{q}}+\cos\frac{q_{x}}{2}\cos\frac{q_{y}}{2}+3\right)}
{\cos\frac{q_{x}}{2}\left(4(D_{\mathbf{q}}+3)\cos\frac{q_{y}-q_{z}}{4}+2\cos\frac{3q_{y}+q_{z}}{4}+\cos\frac{q_{y}+3q_{z}}{4}\right)+4\cos\frac{q_{y}+q_{z}}{4}\left(3D_{\mathbf{q}}+\cos\frac{q_{y}}{2}\cos\frac{q_{z}}{2}+3\right)-\sin\frac{q_{x}}{2}\sin\frac{q_{y}+3q_{z}}{4}+\cos\frac{2q_{x}-q_{y}-3q_{z}}{4}}\\
\frac{2\left(D_{\mathbf{q}}\sin\frac{q_{x}-q_{y}}{4}-\sin\frac{q_{z}}{2}\cos\frac{q_{x}+q_{y}}{4}\right)}{\sin\frac{q_{x}-q_{y}}{2}+\sin\frac{q_{x}-q_{z}}{2}-\sin\frac{q_{y}+q_{z}}{2}}\\
1
\end{array}
\right),
\end{eqnarray}
\end{widetext}
where 
$D_{\mathbf{q}}$ is given in Eq.~(\ref{08}).
Note that these eigenvectors are not normalized
(in contrast to the eigenvectors at $\mathbf{q}=\mathbf{0}$ in Eq.~(\ref{10})).
The corresponding eigenvalues are given in Eqs.~(\ref{07}) and (\ref{09}).
In the limit $\mathbf{q}\to\mathbf{0}$, 
Eq.~(\ref{b04}) 
transforms into $\vert 4{\bf{0}}\rangle$ in Eq.~(\ref{10})
whereas
Eqs.~(\ref{b01}) -- (\ref{b03}) 
yield a linear combination of  $\vert 1{\bf{0}}\rangle$, $\vert 2{\bf{0}}\rangle$, $\vert 3{\bf{0}}\rangle$ given in Eq.~(\ref{10})
depending on the chosen  path along which the limit $\mathbf{q}\to\mathbf{0}$ was taken.

\end{document}